%% file: FabDreamer.tex
\documentclass[sigconf]{acmart}
\usepackage{multirow}
\usepackage{soul}

\newif\ifmarkup
\markupfalse 
\definecolor{markupAdd}{RGB}{217,95,2} 
\ifmarkup
  \newcommand{\add}[1]{{\color{markupAdd}#1}}
  \newcommand{\del}[1]{{\color{red}\st{#1}}}
\else
  \newcommand{\add}[1]{#1}
  \newcommand{\del}[1]{}
\fi

\definecolor{stageDec}{RGB}{24,174,78} 
\definecolor{stageCre}{RGB}{0,144,245} 
\definecolor{stagePre}{RGB}{89,92,237} 

\definecolor{stageDecTint}{RGB}{197,235,211} 
\definecolor{stageCreTint}{RGB}{191,227,252} 
\definecolor{stagePreTint}{RGB}{213,214,250} 
\newcommand{\boxD}[1]{{\setlength{\fboxsep}{1.5pt}\colorbox{stageDecTint}{\textcolor{black}{#1}}}}
\newcommand{\boxC}[1]{{\setlength{\fboxsep}{1.5pt}\colorbox{stageCreTint}{\textcolor{black}{#1}}}}
\newcommand{\boxP}[1]{{\setlength{\fboxsep}{1.5pt}\colorbox{stagePreTint}{\textcolor{black}{#1}}}}
\newcommand{\chal}[1]{\ul{#1}}

\makeatletter
\@ifundefined{extracolsep}{\def\extracolsep{}}{}
\makeatother
\AtBeginDocument{%
  }

\copyrightyear{2026}
\acmYear{2026}
\setcopyright{cc}
\setcctype{by}
\acmConference[UIST '26]{The 39th Annual ACM Symposium on User Interface Software and Technology}{November 02--05, 2026}{Detroit, MI, USA}
\acmBooktitle{The 39th Annual ACM Symposium on User Interface Software and Technology (UIST '26), November 02--05, 2026, Detroit, MI, USA}
\acmDOI{10.1145/3830398.3830571}
\acmISBN{979-8-4007-2856-3/2026/11}

\begin{document}

\title{FabDreamer: Exploring the Image-to-Physical Workflow Through AI-Assisted Layered Fabrication}

\newif\iftapsdept
\tapsdepttrue
\newcommand{\tdept}[1]{\iftapsdept\department{#1}\fi}

\author{Chenfeng Gao}
\authornote{Both authors contributed equally to this research.}
\affiliation{%
  \institution{Northwestern University}
  \city{Evanston}
  \state{Illinois}
  \country{USA}}
\email{chenfenggao2029@u.northwestern.edu}

\author{Zeya Chen}
\authornotemark[1]
\affiliation{%
  \tdept{Institute of Design (ID)}
  \institution{Illinois Institute of Technology}
  \city{Chicago}
  \state{Illinois}
  \country{USA}}
\email{zchen103@hawk.illinoistech.edu}

\author{Anjie Yang}
\affiliation{%
  \institution{University of Ottawa}
  \city{Ottawa}
  \state{Ontario}
  \country{Canada}}
\email{ayang072@uottawa.ca}

\author{Karan Ahuja}
\affiliation{%
  \institution{Northwestern University}
  \city{Evanston}
  \state{Illinois}
  \country{USA}}
\email{kahuja@northwestern.edu}

\author{Danli Luo}
\affiliation{%
  \institution{University of Washington}
  \city{Seattle}
  \state{Washington}
  \country{USA}}
\email{danlil@uw.edu}

\renewcommand{\shortauthors}{Gao, Chen, et al.}

\input{Sections/0-abstract}



\begin{teaserfigure}
  \includegraphics[width=\textwidth]{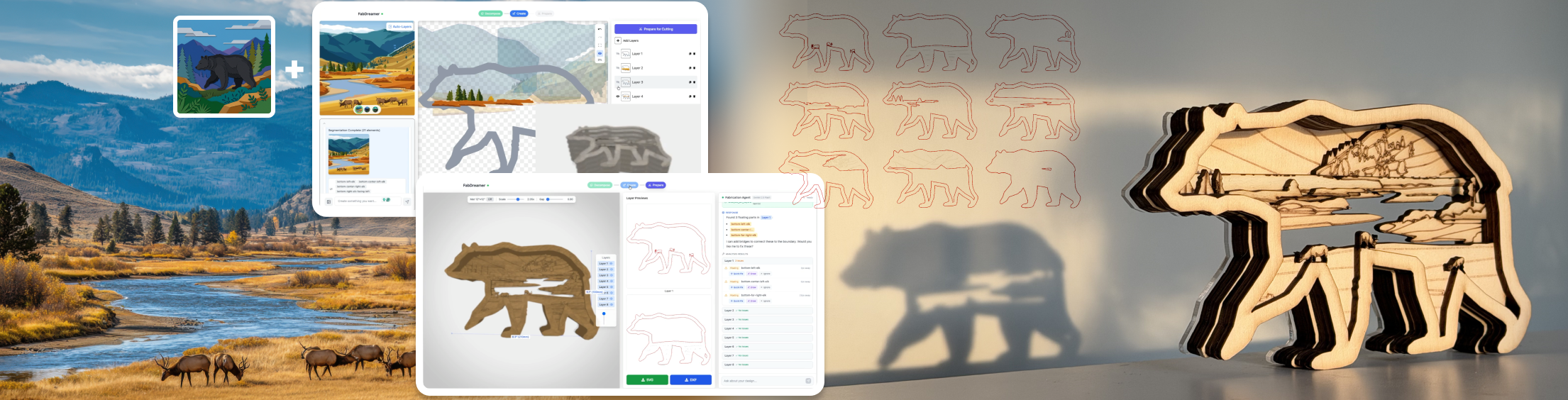}
  \caption{FabDreamer transforms images into physical layered laser-cut artifacts.(Left)~Input: a Yellowstone landscape photograph and a generated bear image. (Center)~Through FabDreamer, the user decomposes, edits, and verifies the layered design with a bear-shaped frame. (Right)~The 
  fabricated eight-layer basswood shadow box.}
  \Description{Teaser figure: a Yellowstone photograph and bear image as inputs, the FabDreamer interface showing decomposition and 3D preview, and the final fabricated layered basswood shadow box.}
  \label{fig:teaser}
\end{teaserfigure}



\maketitle

\clearpage

\input{Sections/1-introduction}


\input{Sections/3-relatedwork}

\input{Sections/4-formative}

\input{Sections/5-system}

\input{Sections/6-study}

\input{Sections/7-findings}

\input{Sections/8-discussion}

\input{Sections/9-conclusion}

\begin{acks}
We thank all study participants across the three rounds of this work for their time, creativity, and candid feedback, and the professional publisher for sharing his practice and expertise. We thank Prof.\ Dr.\ Elizabeth Gerber for her insightful feedback on this project, and the anonymous reviewers, whose suggestions strengthened this paper.
\end{acks}

\bibliographystyle{ACM-Reference-Format}
\bibliography{reference}

\appendix
\input{Sections/appendix}
\end{document}

%% file: Sections/0-abstract.tex

\begin{abstract}

Generative AI lets anyone create rich visual content in seconds, yet translating that content into a physically fabricable artifact still demands manual decomposition, occlusion repair, and structural verification that most tools leave entirely to the user.
\add{We present FabDreamer, an image-to-physical system that carries an image to fabrication-ready SVGs through three stages with deliberately staged AI initiative: (1)~AI leads decomposition into depth-ordered layers, (2)~assists on demand during creative editing with real-time 3D preview, and (3)~advises on structural integrity before export.}
We instantiate this workflow for layered laser-cut art and evaluate it through three rounds including a formative analysis, an early prototype user evaluation (N=13), and a cross-domain practitioner study with specialists from 6 fabrication domains (N=6).
\add{Our findings show that physical awareness during design opens creative opportunities beyond error prevention, that practitioners appropriate the system's generic geometric operations for their own domains, and that the fabrication agent covers geometry-readable constraints while domain knowledge remains with the maker.}

\end{abstract}

%% file: Sections/1-introduction.tex

\section{Introduction}
\label{sec:intro}

Generative AI has transformed visual content creation.
A text prompt can produce a detailed illustration in seconds, and novices and experts use these tools to explore ideas, iterate on compositions, and accelerate workflows that once required hours of manual work~\cite{openai_dalle, midjourney}. 
But generating an image is only the beginning when the goal is a physical artifact. 
Turning a flat image into a layered laser-cut shadow box, a vinyl sticker scene, a paper-cut card, or an architectural landscape model requires decomposing the image into physically separable parts, reconstructing content hidden behind foreground elements, and ensuring every piece survives the fabrication process. 
This image-to-physical transition is where the workflow breaks down.

The individual capabilities this transition requires already exist: segmentation models isolate objects~\cite{ravi2024sam2}, generative models reconstruct occluded content, and constraint checkers flag structural problems~\cite{sethapakdi2021fabricaide}.
\add{What remains unsolved is their integration. These capabilities live in separate tools, each optimized for screen-based output, and the knowledge required to carry a design across them under physical constraints is tacit, domain-specific, and built through years of material experience~\cite{twigg2021tools, sethapakdi2021fabricaide}. Layered laser-cut art illustrates the resulting burden: an active maker community sustains the craft, yet getting started demands illustration, digital layering, and laser setup simultaneously, and practitioners must manually maintain fabrication constraints across every tool boundary. Consequently, the community treats preparation rather than creation as the principal bottleneck.}

HCI research has made substantial progress on constraint-aware fabrication tools~\cite{sethapakdi2021fabricaide, leake2024scrapmap} and on AI-assisted 3D printing and shape generation~\cite{faruqi2023style2fab, qian2024shape}. 
However, these systems either assume a prepared design file as input or operate on 3D models rather than images.
A pipeline that carries an arbitrary image to a physically fabricable artifact thus remains both a practical gap and a research opportunity: it raises the question of how much initiative AI should take at each step of a workflow whose errors, once cut, are irreversible.

We present \textbf{FabDreamer}, an AI-assisted system that addresses this question by deliberately calibrating human-AI initiative across three stages according to the recoverability of errors at each point: an erroneous decomposition can simply be regenerated, whereas a cut cannot be undone.
In \boxD{Stage~1 (Decompose)}, AI proposes a depth-ordered layer decomposition that users \add{iterate on and revise} before execution. In \boxC{Stage~2 (Create)}, users lead all creative decisions on a canvas with real-time 3D preview; AI assists only on demand. In \boxP{Stage~3 (Prepare)}, an AI fabrication agent detects structural issues (floating elements, thin features) and the user decides how to resolve each one.

\add{We instantiate this workflow for layered laser-cut art, a domain that exercises the full image-to-physical pipeline and whose core geometric constraints recur across other cutting and layered fabrication domains~\cite{sethapakdi2021fabricaide}. We developed FabDreamer through three rounds (Sec.~\ref{sec:formative}): a formative analysis of practitioner tutorials and an expert interview, an early prototype study (10 novices, 3 experts), and a cross-domain study in which 6 practitioners applied FabDreamer to their own practice (Sec.~\ref{sec:evaluation}). The domains participants brought extend well beyond visual content creation: architectural site models, tactile graphics for blind and low-vision readers, actuator prototypes, and pop-up structures, fabrication tasks in which the image is the starting point of a functional physical object rather than the end product.}

\add{We contribute: (1) \textbf{FabDreamer}, an image-to-physical system for layered 2.5D art, combining hybrid segmentation routing, generative occlusion reconstruction, and a fabrication agent that verifies structural integrity before export, with AI initiative staged across the workflow: leading decomposition, assisting creation, and advising preparation. The design targets four persistent challenges that our formative study identified in current practice (Sec.~\ref{sec:formative}). (2) \textbf{Empirical findings} from multi-round development and a cross-domain practitioner study that offer guidance for future AI-assisted fabrication tools: how practitioners adapt the staged workflow to their own domains, how physical awareness during design creates creative opportunities beyond error prevention, and where the boundary lies between constraints an agent can detect from geometry and knowledge that only practitioners supply.}

%% file: Sections/3-relatedwork.tex

\section{Related Work}
\label{sec:relatedwork}
 
We situate FabDreamer in four research areas: image-to-physical fabrication, creativity support in fabrication, generative AI for visual and physical design, and human-AI initiative in creative workflows.
 
\subsection{Image-to-Physical Fabrication}
\label{subsec:rw-fabrication}
 
Personal fabrication follows a common workflow of design, prepare, fabricate, and post-process across a wide range of machines and materials~\cite{baudisch2017personal, gershenfeld2005fab, gershenfeld2012make}, a structure that recurs across the 14 fabrication workflows \citet{feng2026cameleon} implemented. This consistency implies that advances at any stage can in principle transfer across domains, yet the stages are typically addressed by separate tools with no shared representation, leaving practitioners to bridge the gaps manually.
 
Within laser cutting, HCI research has focused primarily on functional prototyping: assembling 3D objects from flat stock~\cite{baudisch2019kyub, mueller2013laserorigami, nisser2021laserfactory}, fabricating from stacked layers~\cite{umapathi2015laserstacker}, and translating 3D models into 2D cutting layouts~\cite{beyer2015platener, park2022foolproofjoint, roumen2021autoassembler}.
However, tools for artistic compositions are under-explored, despite a rich cultural history in traditions such as Chinese \textit{jianzhi}~\cite{bodolec2012chinese} and growing commercial demand on maker platforms. These practices rely on spatial arrangement of discrete material layers, creating depth through stacking. They remain largely manual because the required knowledge is rarely documented, accumulating instead within individual practice through trial and error.

Prior work on constraint-aware design tools shows that surfacing this knowledge early in the design process improves outcomes.
Fabricaide~\cite{sethapakdi2021fabricaide} demonstrated this principle for 2D cutting machines by flagging material and structural issues during design; ScrapMap~\cite{leake2024scrapmap} and Towards Zero-Waste Furniture~\cite{koo2016towards} apply similar constraint-as-information logic to quilting and furniture, respectively. However, all three systems assume a \textit{prepared design file} as input and optimize existing geometry rather than supporting the creative process that produces it. 
FabDreamer addresses an upstream gap from an arbitrary image to a fabrication-ready layered design, with constraints informing creative decisions throughout rather than gating the output at the end.
 
\subsection{Creativity Support in Fabrication}
\label{subsec:rw-cst-power}
 
End-to-end fabrication pipelines risk disempowering users by hiding the abstractions that structure their work. \citet{li2023beyond} formalize this as the \textit{normative ground} of creativity support tools (CSTs): the implicit constraints a tool imposes on how practitioners think, act, and express themselves. 
They argue that mitigating this power imbalance requires \textit{vertical movement} (letting users inspect, modify, and move between levels of abstraction) and \textit{horizontal movement} (letting users compose tools through interoperable representations). 
Their analysis of digital fabrication pipelines specifically warns against all-in-one systems that prevent users from working with lower-level primitives or integrating with other tools.
 
Empirical observations of maker communities reinforce this argument. \citet{twigg2021tools} found that \#PlotterTwitter practitioners preferred composing lightweight, self-authored scripts over monolithic tools, sharing components that others could integrate into their own workflows. 
\citet{devendorf2015being} showed that how agency is distributed in digitally-mediated making fundamentally shapes the maker's experience. 
\citet{jacobs2018extending} demonstrated that computational tools designed around practitioner workflows enhance practice without displacing expertise. 
In paper cutting specifically, computational approaches range from algorithmic stylization~\cite{meng2010artistic} to ideation while preserving cultural connotation ~\cite{wang2025harmonycut}.
These works demonstrate that AI can serve creative exploration in traditional cutting arts without overriding practitioner intent.
 
These findings suggest that effective image-to-physical tools must support both vertical movement (inspecting and modifying AI abstractions) and horizontal movement (interoperating with existing tools through standard formats), while keeping the normative ground of what constitutes a ``good'' design with the practitioner.
 
\subsection{Generative AI for Visual and Physical Design}
\label{subsec:rw-genai}
 
Generative AI can now produce visual content (text-to-image models like DALL\textperiodcentered E~\cite{openai_dalle}, Midjourney~\cite{midjourney}, and Stable Diffusion~\cite{rombach2021highresolution}), decompose images into masks (SAM2~\cite{ravi2024sam2}), and even separate images into editable layers (Qwen-Image-Layered~\cite{yin2025qwenimagelayered}, Canva Magic Layers, Collaposer~\cite{zhou2026collaposer}). But all of these systems are optimized for screen-based output: their masks are visually accurate but not physically separable, and their compositions are screen-ready but not fabrication-ready.
 
In the 3D domain, AI has been applied more directly to physical making. Style2Fab~\cite{faruqi2023style2fab} introduces functionality-aware segmentation for 3D printing, SHAPE-IT~\cite{qian2024shape} uses LLMs for shape-display control, and \citet{liu20233dall} and \citet{shen2024neural} integrate generative models into 3D design workflows. A key lesson from this line of work is that AI-generated geometry frequently fails physical constraints: MechStyle~\cite{faruqi2025mechstyle} found that only 26\% of AI-modified 3D models remain structurally viable. This underscores why constraints must be integrated \textit{throughout} the pipeline rather than checked after the fact.
 
For 2D fabrication, the gap is wider. Text-to-SVG systems like VectorFusion~\cite{jain2023vectorfusion} and SVGDreamer~\cite{xing2024svgdreamer} produce editable vectors but are disconnected from fabrication workflows. Commercial laser-cutting platforms (xTool AIMake, Glowforge Magic Canvas~\cite{glowforge}) generate single-layer designs ready for cutting but skip creative editing entirely, offering no control over decomposition or composition. The gap thus remains open at both ends: decomposition tools stop at screen-ready masks, while fabrication platforms omit the creative editing between image and artifact. By connecting decomposition, creative editing, and structural verification in a single pipeline, FabDreamer closes this gap and provides a setting for examining how practitioners distribute work between themselves and AI when the output is physical.
 
\subsection{Human-AI Initiative in Creative Workflows}
\label{subsec:rw-initiative}
 
Human-AI co-creation research consistently identifies user \textit{control} as the primary determinant of how users relate to AI-assisted output, outweighing the amount of assistance provided. \citet{draxler2024ghostwriter} showed this for writing; \citet{gero2023social} found that creative workers prefer on-demand over unsolicited AI assistance; and \citet{lee2022coauthor} observed that higher AI-suggestion acceptance correlates with reduced engagement. Effective AI assistance must therefore be calibrated to the task: proactive where users want help, restrained where they want agency. Frameworks like COFI~\cite{rezwana2023cofi} and MOSAAIC~\cite{issak2025mosaaic} operationalize this insight by modeling initiative and authority as tunable dimensions, and \citet{singh2025systematic} confirms that AI should lead during ideation but step back during evaluation. \add{Recent comparisons of oversight strategies for computer-use agents further show that the appropriate degree of human checking depends on how reversible the agent's actions are~\cite{chen2026comparing}.}
 
However, these frameworks have been validated primarily in screen-based contexts such as writing and drawing. When the output is a physical artifact, misaligned AI initiative carries higher stakes: a segmentation error that is cosmetic on screen becomes a structural failure when laser-cut. \citet{zoran2013hybrid} found that hybrid artifacts combining digital fabrication with handwork maintain uniqueness through material engagement, and \citet{campbell2025effort} provides evidence that effort investment increases the maker's perceived meaning of the artifact. No existing framework, however, has been tested in fabrication contexts where AI errors carry irreversible material consequences.

FabDreamer addresses this gap through an iteratively developed pipeline whose design was shaped by three rounds of evaluation, described next.

%% file: Sections/4-formative.tex

\section{Iterative Design Process}
\label{sec:formative}

Designing a system at the intersection of these challenges required building and testing alongside practitioners, because the tacit knowledge governing image-to-physical workflows is not fully captured in prior literature~\cite{twigg2021tools, feng2026cameleon}. \add{Within layered laser-cut art, chosen as our instantiation domain for the reasons given in Section~\ref{sec:intro},} we structured the design process as three rounds, each directly informing the next:

\textbf{Round 1: Formative Study.}
\add{We analyzed 25 YouTube tutorials on multilayered laser-cut design (2019--2024), selected by topical criteria rather than popularity, together with a 60-minute semi-structured interview with a professional publisher with over 10 years of experience in layered laser-cut production (selection and coding detailed in Appendix~\ref{appendix:formative}). The analysis revealed four persistent challenges that became design targets for the pipeline and evaluation criteria for later rounds; the quantitative patterns below derive from the tutorial corpus, and the practitioner accounts in \chal{C2--C4} derive from the interview.}

\textit{\chal{C1: Element Sourcing Constraints.}}
76\% of tutorials relied on existing vector libraries rather than personal images, because 
isolating and preparing arbitrary images for fabrication required manual tracing, format 
conversion, and prior knowledge of what would segment cleanly. This effectively constrained 
creative input to what was already available in vector form.

\textit{\chal{C2: Preparation Bottlenecks.}}
Practitioners identified preparation as the most time-intensive phase\add{---``getting clean edges while preserving detail is always a balance,'' as the publisher put it}. Separating a flat
image into discrete cuttable layers required manual masking of each element and reconstruction
of content occluded by foreground objects, work that was largely invisible in finished
tutorials and routinely underestimated by novices.

\textit{\chal{C3: Limited Spatial Awareness.}}
Only 2 of 25 tutorials applied 3D visualization to preview the designs. Practitioners relied on mental
simulation to reason about how flat layers would read as a stacked physical artifact, a skill
built through years of experience. \add{The publisher, who also teaches layered design, watched novices hit this wall repeatedly: ``when my kids are doing it in Adobe Illustrator, they can't visualize the spatial relationship... I have to lead them layer by layer.''} Misaligned elements and misjudged depth relationships were
commonly discovered only after cutting.

\textit{\chal{C4: Invisible Fabrication Constraints.}}
Material-specific limits, such as minimum feature widths and elements with no physical 
connection to the board, were rarely documented and almost never surfaced during 
design~\cite{faruqi2025mechstyle, sethapakdi2021fabricaide}. Practitioners reported 
discovering these constraints exclusively through failed cuts\add{; with no tool that makes them visible earlier, the publisher's only safeguard was manual gatekeeping: ``every time before my kids cut, I go through the file with my own eyes and fix it for them.''}


\begin{figure}[t]
  \centering
  \includegraphics[width=\linewidth]{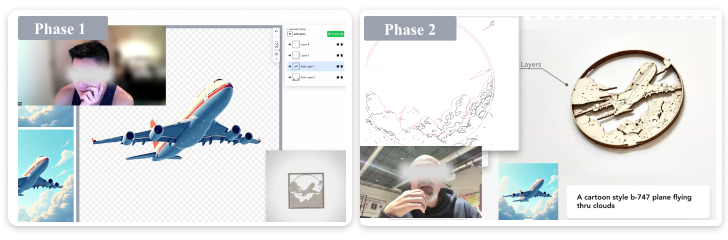}
  \caption{\add{Round~2 study. Left: example of a novice creating a layered design with the early prototype. Right: example of an expert reviewing the fabricated output and design process.}}
  \label{fig:early-prototype}
  \Description{Two-phase pilot study. Phase 1 (left): novice participant using the early prototype, with an example input photo, the decomposed layers in the tool, and the 3D preview. Phase 2 (right): expert reviewing the output, with the SVG export and the fabricated laser-cut artifact.}
\end{figure}

\textbf{Round 2: Early Prototype Study.} Informed by these challenges, we built an early prototype with AI content generation, automated layer decomposition with occlusion reconstruction, a real-time 3D preview, and SVG export, and evaluated it with 10 novices creating designs and 3 domain experts reviewing the fabricated outputs (Figure~\ref{fig:early-prototype}; detailed procedure in Appendix~\ref{appendix:pilot}). All 10 novices successfully created 3 to 5 layer designs that survived laser cutting and physical assembly in approximately 25 minutes. The 3D preview received the highest satisfaction rating (mean = 4.9/5), confirming that the pipeline can address \chal{C1}~(sourcing) through \chal{C3}~(spatial awareness).

Two observations from the pilot further informed our design. First, packing layer detection and segmentation into a single automated pass meant that when the AI misinterpreted spatial relationships, users had no way to intervene or recover. 
Participants needed to \add{\textit{iterate} on} the decomposition, not simply accept or reject it. Second, expert reviewers identified that \chal{C4}~(fabrication constraints) remained entirely unaddressed: the prototype exported SVGs without checking physical viability, missing floating parts, features too thin to survive cutting, and the need to reason about real physical dimensions at fabrication scale.
These observations, together with human-AI collaboration frameworks recommending stage-appropriate AI roles~\cite{gero2023social, rezwana2023cofi, singh2025systematic}, led to FabDreamer's three-stage design with varying AI initiative.

\boxD{Stage~1 (Decompose)} addresses \chal{C1} (sourcing) and \chal{C2} (preparation): AI proposes a layer decomposition that users \add{revise} and confirm before segmentation executes, replacing the single automated pass that left no room for intervention. \boxC{Stage~2 (Create)} addresses \chal{C3}~(spatial awareness): the user leads all creative decisions on a canvas with real-time 3D preview; AI assists only on explicit request. \boxP{Stage~3 (Prepare)} addresses \chal{C4}~(fabrication constraints): an AI fabrication agent detects structural issues (floating elements, thin features) and explains their physical consequences, but the user decides whether and how to resolve each issue.



\textbf{Round 3: Practitioner Study.} A cross-domain study evaluated FabDreamer with 6 fabrication practitioners spanning the full spectrum of laser-cut experience, from a fabric artist with no prior laser-cut exposure to a makerspace manager with extensive expertise, to (1) validate FabDreamer on a broader and more diverse user base than Round 2 and (2) investigate whether the three-stage workflow generalizes beyond layered laser-cut art to other fabrication practices. Round 3 is detailed in Section~\ref{sec:evaluation}.

%% file: Sections/5-system.tex

\section{FabDreamer}
\label{sec:system}

\add{FabDreamer carries any image to a fabrication-ready SVG. This section describes what the user experiences at each stage,} using the running example in Figure~\ref{fig:teaser}: turning a Yellowstone landscape photograph into a layered shadow box with a bear-shaped frame.

\subsection{\texorpdfstring{\boxD{Stage 1: Decompose}}{Stage 1: Decompose}}
\label{subsec:decompose}

Rather than automating decomposition in a single pass, Stage~1 has the AI propose a structured layer decomposition that the user reviews and reshapes through natural language before segmentation begins. 
In the running example (Figure~\ref{fig:teaser}), the user uploads a Yellowstone photograph, simplifies the distant mountains through a natural-language update, then \add{revises} the proposed eight-layer structure until it matches the intended depth arrangement.

\begin{figure}[t]
  \centering
  \includegraphics[width=\linewidth]{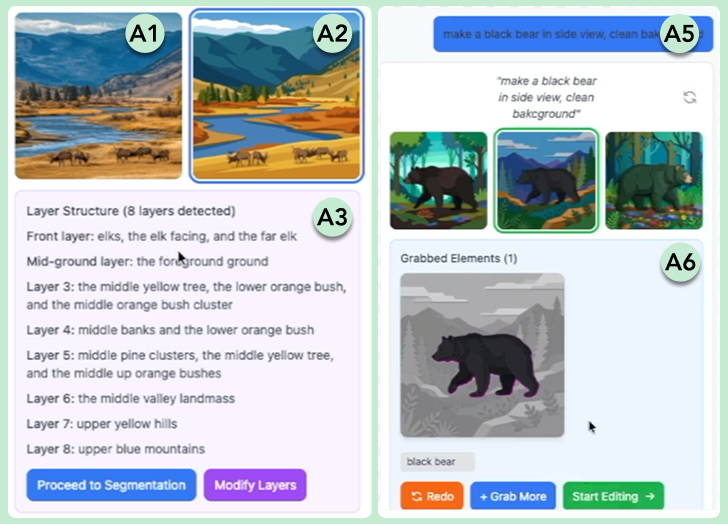}
  \caption{Stage~1: Decompose. (A1)~Uploaded photograph. (A2)~Simplified version via natural-language update. (A3)~Layer proposal with user modification. (A5)~Text-to-image generation produces three candidates. (A6)~Grab extracts the selected element.}
  \label{fig:decompose}
  \Description{Single-column figure showing Stage 1. Left column: original Yellowstone photo (A1), simplified version (A2), and eight-layer proposal with user modification input and Modify Layers button (A3). Right column: three generated bear candidates from text prompt (A5) and the grabbed bear element ready for editing (A6).}
\end{figure}

\subsubsection{\textbf{Image Input and Refinement}}

The user provides an image by uploading a photograph (Figure~\ref{fig:decompose}, A1) or generating three candidates from a text prompt (Figure~\ref{fig:decompose}, A5). Every image supports an \textit{update} command for refinement through natural language---stylizing, adding or removing elements, or adjusting composition (Figure~\ref{fig:decompose}, A2).
Each update \add{regenerates the whole image (a property of current image-editing models, so untouched regions may drift slightly) and is kept as} a new version\add{;} all variants can be independently analyzed in parallel.

\subsubsection{\textbf{Layer Proposal and Extraction}}

\add{Round~2's single automated pass left users no way to intervene when the AI misread the scene (Sec.~\ref{sec:formative}),} so the system proposes a depth-ordered layer structure in natural language (Figure~\ref{fig:decompose}, A3) first. The user can confirm to proceed, or modify through language instructions (e.g., ``move the cloud to layer~2''). Because the VLM analyzes the image during modification, users can also request elements the initial analysis missed (e.g., ``also add the left river bank''). Only after confirmation does segmentation begin.

\begin{figure}[t]
  \centering
  \includegraphics[width=\linewidth]{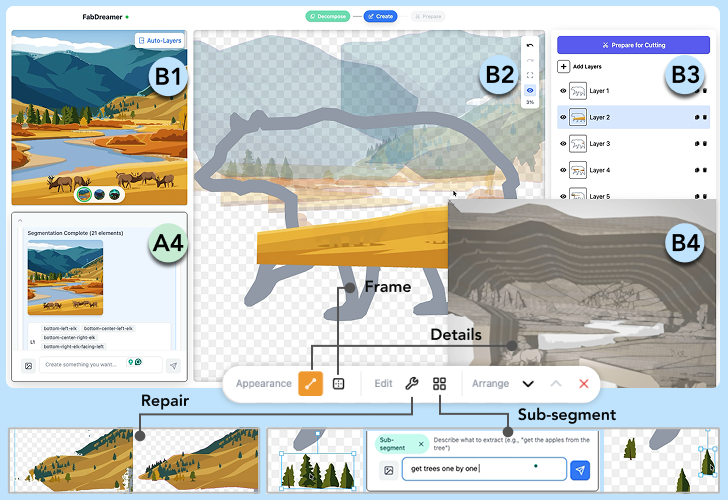}
  \caption{Stage~2: Create. (A4)~Segmented elements ready for editing. (B1)~Source image and segment pool. (B2)~2D canvas with bear-shaped frame and see-through layer overlay. (B3)~Layer management. (B4)~Real-time 3D stacked preview. The toolbar for each element provides options to convert to frame, show/hide inner details, repair noisy edges, and further segment the element.}
  \label{fig:create}
  \Description{Single-column figure showing Stage 2 Create workspace. Top-left: source image with Auto-Layers button (B1). Bottom-left: segmentation complete panel with 21 elements (A4). Center: 2D canvas with bear frame overlay and see-through editing (B2). Top-right: layer management panel (B3). Bottom-right: 3D stacked preview (B4).}
\end{figure}

Elements then appear progressively as the pipeline processes each layer, approximately one minute per layer (Figure~\ref{fig:create}, A4). If segmentation quality is insufficient, the user can redo a single layer or from any layer onward (cascade), preserving previously extracted results. The decomposition pipeline is detailed in Sec.~\ref{subsec:decomposition}. Alternatively, the \textit{Grab} pathway skips full decomposition: the user describes a target element, the system generates candidates and extracts the selected one with automatic occlusion handling (Figure~\ref{fig:decompose}, A5--A6).

\subsection{\texorpdfstring{\boxC{Stage 2: Create}}{Stage 2: Create}}
\label{subsec:create}

With decomposition complete, users compose elements into a spatial design on a canvas with real-time 3D preview. The user leads all creative decisions; AI assists only on explicit request. In the running example, the user repairs rough edges from segmentation, toggles engrave detail while checking the 3D preview, generates a bear silhouette to use as a custom frame, and splits a dense tree cluster into individual trees through sub-segmentation.

\subsubsection{\textbf{Canvas and 3D Preview}}

Elements from Stage~1 appear in a segment pool (Figure~\ref{fig:create}, B1). An Auto Layout places all elements onto their corresponding layer canvases; from there, users reposition, scale, rotate, and reassign elements between layers (Figure~\ref{fig:create}, B2). Each segment can display as \textit{traced} (internal detail lines for engraving) or \textit{outline} (cut path only), and any segment can be converted into a frame---a ring-shaped cutout serving as the physical layer boundary with adjustable border width. A real-time 3D preview (Figure~\ref{fig:create}, B4) extrudes each layer's SVG into stacked geometry with material textures, updating continuously as users edit, so they can check spatial relationships throughout the process rather than only at the end.

\subsubsection{\textbf{On-Demand AI}}

On-demand AI tools include sub-segmentation via chat (``separate the head from the body''), Grab for importing elements from additional images (Sec.~\ref{subsec:decompose}), and repair for noisy segments. Repair regenerates a clean version without re-running the full pipeline; sub-segmentation splits a compound element into individually placeable parts. Both are detailed in Sec.~\ref{subsec:decomposition}.

\subsection{\texorpdfstring{\boxP{Stage 3: Prepare}}{Stage 3: Prepare}}
\label{subsec:prepare}

A thin branch looks fine on screen but crumbles under the laser; a floating cloud has no physical connection to the board. Stage~3 makes these structural issues visible before cutting through an AI fabrication agent that checks geometry-level constraints (floating elements, thin features) detectable from the SVG alone. The user always decides how to resolve each issue. In the running example, the user sees floating bush pieces flagged by the agent, Quick Fixes two with bridge previews, leaves the third floating for manual gluing, switches a layer to acrylic, and exports nine SVGs ready for cutting.

\begin{figure}[t]
  \centering
  \includegraphics[width=\linewidth]{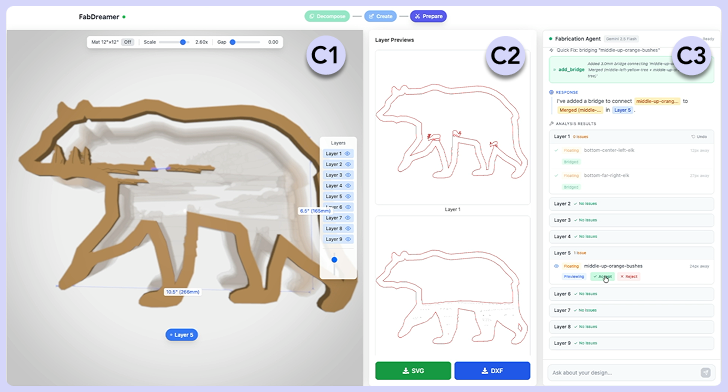}
  \caption{Stage~3: Prepare. (C1)~3D preview at physical scale with layer toggles. (C2)~Per-layer SVG previews with cut (red) and engrave (black) paths. (C3)~Fabrication agent reporting floating parts with resolution options.}
  \label{fig:prepare}
  \Description{Single-column figure showing Stage 3 Prepare workspace. Left: 3D preview at 210mm scale with 8 layer toggles (C1). Center: per-layer SVG previews showing cut and engrave paths (C2). Right: fabrication agent reporting 3 floating elk with Quick Fix, Draw, and Ignore buttons (C3).}
\end{figure}

\begin{figure*}[t]
  \centering
  \includegraphics[width=\textwidth]{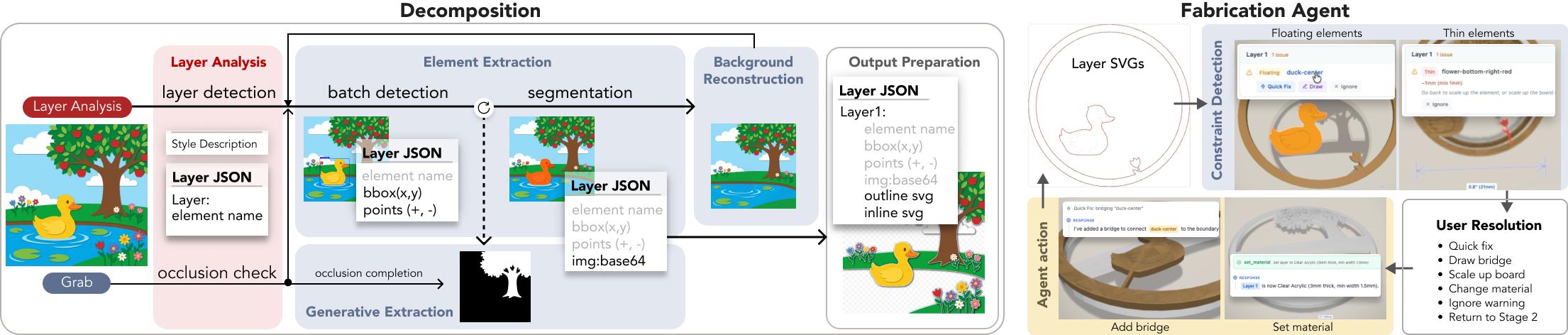}
  \caption{Backend architecture. \textbf{Left:} Decomposition pipeline processes layers front to back: analyze, detect, segment, extract, inpaint, repeat. \textbf{Right:} Fabrication agent loop: constraint detection (floating elements, thin elements) surfaces issues, the user chooses a resolution, the agent executes the action (add bridge, set material), and re-checks the modified layer.}
  \label{fig:architecture}
  \Description{Full-width architecture diagram. Left two-thirds: decomposition pipeline showing layer analysis, batch detection, SAM2 segmentation with quality gate, element extraction, SVG conversion, and style-matched inpainting, with filled image feeding back for the next layer. Right one-third: fabrication agent loop showing constraint detection (floating and thin elements), user resolution options, and agent actions (add bridge, set material), with a re-check arrow closing the loop.}
\end{figure*}

\subsubsection{\textbf{Fabrication Agent Interaction}}

Entering Prepare mode triggers the fabrication agent's analysis loop (Figure~\ref{fig:prepare}, C3). The agent detects constraint violations, highlights them in the 3D view, and presents resolution options: auto-bridge, user-drawn bridge, ignore, or return to Stage~2. The agent re-checks after each action. Resolution is always the user's decision because the agent detects geometric problems but lacks the domain knowledge to decide how they should be resolved. A user may draw a bridge to a specific neighboring element rather than the nearest anchor for visual coherence, or intentionally leave an element floating to cut it as a separate piece for manual attachment after fabrication. The scale slider (Figure~\ref{fig:prepare}, C1) sets the physical size of the artifact; changing it triggers re-evaluation since features that survive at one scale may be too thin at another. The detection algorithms and agent tools are detailed in Sec.~\ref{subsec:fabagent}.

\subsubsection{\textbf{Export}}

Per-layer SVG previews show cut paths (red) and engrave paths (black) side by side (Figure~\ref{fig:prepare}, C2), giving users a final check before export. The system exports one SVG per layer along with corresponding PNG files, packaged as a ZIP file for direct import into fabrication software such as xTool Creative Space, LightBurn, or Adobe Illustrator.


\section{Implementation}
\label{sec:implementation}

FabDreamer's backend consists of two components (Figure~\ref{fig:architecture}): a \textit{decomposition pipeline} that extracts and reconstructs elements layer by layer, and a \textit{fabrication agent} that checks each layer for floating elements and thin features, presents issues to the user with resolution options, and re-checks after each fix. The decomposition pipeline proceeds in four phases (layer analysis, element extraction, background reconstruction, output preparation), each passing its result to the next.

We implement the system as a web application with a React and Three.js frontend and a FastAPI backend. All images are processed at $1024 \times 1024$ pixels, mapping to a $304.8 \times 304.8$\,mm (12'' $\times$ 12'') physical board. The system uses three categories of AI models: a VLM (Gemini~3.1 Pro) for image analysis, detection, and scene understanding; an image generation model (Gemini Flash) for stylization, inpainting, and reconstruction; and SAM2 Large~\cite{ravi2024sam2} for pixel-level segmentation, self-hosted on a Mac Mini (M4) with MPS acceleration. Stage~3's fabrication agent uses Gemini~2.5 Flash with function calling. Per-component model versions, prompt design patterns, and engineering methodology are detailed in Appendix~\ref{appendix:llm}.

\subsection{Decomposition Pipeline}
\label{subsec:decomposition}

The decomposition pipeline transforms a confirmed layer proposal into individually editable elements with vector cut and engrave paths, in the four phases shown in Figure~\ref{fig:architecture} (left).


\subsubsection{\textbf{Layer Analysis}}


Given the confirmed layer proposal from Stage~1 (Sec.~\ref{subsec:decompose}), a VLM call analyzes the source image and returns a structured list of elements grouped by depth layer, each with a natural-language description using absolute spatial references (Appendix~\ref{appendix:llm}). A separate call produces a one-time style description covering art style, spatial layout, background surface, and texture language, stored for later inpainting.

\subsubsection{\textbf{Element Extraction}}

The pipeline processes one layer at a time, front to back. For each layer, a VLM detection call locates every element and returns per-element bounding boxes along with positive and negative point prompts that guide SAM2 toward the correct region. To handle the wide variation in element types---from sharp-edged buildings to diffuse tree canopies---we designed a multi-stage extraction cascade (Figure~\ref{fig:architecture}, left). SAM2~\cite{ravi2024sam2} attempts segmentation first using the bounding box and point prompts. If the resulting mask scores below a confidence threshold or is disproportionate to the bounding box, the system feeds the initial low-resolution output back into SAM2 as a refinement pass. When SAM2 cannot produce an adequate mask---common for elements with structural openings such as tree canopies or rock arches---the system falls back to generative extraction (Figure~\ref{fig:architecture}, differential mask): the image generation model redraws the element on a uniform background, and pixel-wise comparison against the original produces the mask. This cascade prioritizes speed (SAM2 is an order of magnitude faster than generative extraction) while ensuring coverage across element types. Elements with noisy or incomplete masks can be repaired on demand in Stage~2 (Sec.~\ref{subsec:create}).

Occluded elements are handled implicitly by the front-to-back processing order: once a front layer's elements are extracted and inpainted away, the hidden portions of deeper elements are revealed for subsequent detection. The Grab pathway (Sec.~\ref{subsec:decompose}) uses a similar cascade for single-element extraction, with an additional occlusion check: if a VLM determines the target is partially hidden, a generative model draws the complete element on a magenta background, and chroma-key removal produces the mask and SVG outline, bypassing SAM2 entirely.

\subsubsection{\textbf{Background Reconstruction}}

After all elements in a layer are extracted, the pipeline fills the regions they occupied so that the next layer can be detected against a clean image. Elements are removed front-to-back, one layer at a time. Before filling each hole, a VLM looks at the current image, the hole to fill, and the stored image style description, then writes specific fill instructions (e.g., ``extend the red-and-navy plaid fabric pattern at the same 45-degree diagonal direction''). The prompt also lists which elements have already been removed, to prevent the model from accidentally recreating them (Appendix~\ref{appendix:llm}). The filled image becomes input for the next layer. This cycle of extract-then-fill repeats until all layers are processed.

\subsubsection{\textbf{Output Preparation}}

Each extracted element produces two vector representations: an outline SVG (mask boundary converted to a cut path) and an engrave SVG (internal detail lines via Sobel edge detection and Potrace vectorization).

The pipeline also supports two on-demand operations from Stage~2 (Sec.~\ref{subsec:create}). \textit{Repair} uses the same chroma-key isolation as the Grab pathway's occlusion completion: the generative model receives the element name, a cropped region of the original scene (25\% padded bounding box), and the noisy extraction, then regenerates the element on a magenta background (Appendix~\ref{appendix:llm}). \textit{Sub-segmentation} re-runs the batch detection pipeline on a single element's image, treating it as a new scene to decompose into parts.

\subsection{Fabrication Agent}
\label{subsec:fabagent}

\add{The fabrication agent covers the two \chal{C4}~(fabrication constraints) failure modes (Sec.~\ref{sec:formative}) detectable from SVG geometry alone, floating and thin elements. The agent operates in a loop (Figure~\ref{fig:architecture}, right): it detects violations, the user chooses a resolution (Sec.~\ref{subsec:prepare}), and the agent executes the fix and re-checks the layer, until all issues are resolved or the user exports.}

\subsubsection{\textbf{Constraint Detection}}
\label{subsubsec:constraint}

%
Before analysis, the system merges overlapping segments within each layer. A union-find graph treats all segments and the frame ring as nodes; pairwise overlap detection inflates each polygon by 5\,px \add{(in the fixed $1024$\,px image space, independent of board size), turning touching edges, which share zero area and would be missed by a strict boolean intersection, into measurable overlap.} Connected components are merged via boolean union into single outlines, eliminating internal cut lines between touching segments.

The agent then detects two classes of physical constraints (Figure~\ref{fig:architecture}, right). \textit{Floating elements}: any merged component not connected to the frame ring has no physical support in the final artifact. The agent highlights the floating part in the 3D view and reports its distance to the nearest anchor. \textit{Thin elements}: \add{a region is too thin when it is narrower than the material's \emph{minimum survivable width} $w_{\min}$, the narrowest strip that survives cutting without snapping or burning away (plywood $\geq$ 1.0\,mm, acrylic $\geq$ 1.5\,mm, set with the material). Detection is a morphological opening: each polygon is eroded inward by $w_{\min}/2$ and dilated back, and whatever fails to survive the round-trip is flagged with its measured width. The erosion distance is scale-aware: $\mathrm{erosion} = (w_{\min}/2) \times 1024 / (\mathrm{board}_{\mathrm{mm}} \times \mathrm{scale})$.} Changing the fabrication scale triggers re-evaluation, since features that survive at one size may fall below the threshold at another.

\subsubsection{\textbf{Agent Action}}
\label{subsubsec:action}

The agent executes two types of actions based on the user's resolution choice (Sec.~\ref{subsec:prepare}). \textit{Add bridge}: a bridge polygon connects the floating part to its nearest anchor, or the user draws a bridge directly on the 3D view for precise placement. Each bridge is previewed as a translucent ghost overlay before committing. Once committed, the transitive union merge detects the chain (element, bridge, frame) and produces a single continuous outline with no internal cut lines. \textit{Set material}: the agent updates the material preset for a layer through natural language (e.g., ``change the last layer to acrylic''), changing both the 3D preview appearance and the physical thresholds for constraint detection ($w_{\min}$ varies by material).


\subsection{Pipeline Evaluation}
\label{subsec:pipeline-eval}

To validate generalization and identify failure modes, we tested the pipeline on 24~images (12~real photographs, 12~AI-generated) from Vecteezy across six categories. \add{Table~\ref{tab:pipeline-eval} reports which extraction route resolved each element.} The majority of elements resolved through the SAM2 cascade; the most challenging cases were diffuse natural boundaries in real photographs (distant mountains, fog gradients), which often required Stage~2 repair. Each layer took approximately 39\,s, totaling 3--4 minutes per image\add{; the bottleneck is API latency (VLM detection and generative inpainting) rather than local computation}. Detailed extraction statistics and per-category breakdowns are reported in Appendix~\ref{appendix:pipeline-eval}.

\begin{table}[h]
  \caption{\add{Extraction route per element (238~elements across 107~layers). Each element resolves at the first route that produces an adequate mask; the SAM2 first pass resolves at mean confidence 0.91, generative extraction handles structural openings, and the bounding-box fallback covers low-contrast regions.}}
  \label{tab:pipeline-eval}
  \small\setlength{\tabcolsep}{2pt}
  \add{\begin{tabular*}{\columnwidth}{@{\extracolsep{\fill}}lcccc}
    \toprule
    Route & SAM2 first & SAM2 refine & Generative & Bbox \\
    \midrule
    Elements & 77\% & 4\% & 12\% & 7\% \\
    \bottomrule
  \end{tabular*}}
\end{table}


%% file: Sections/6-study.tex

\begin{figure*}[t]
  \centering
  \includegraphics[width=\textwidth]{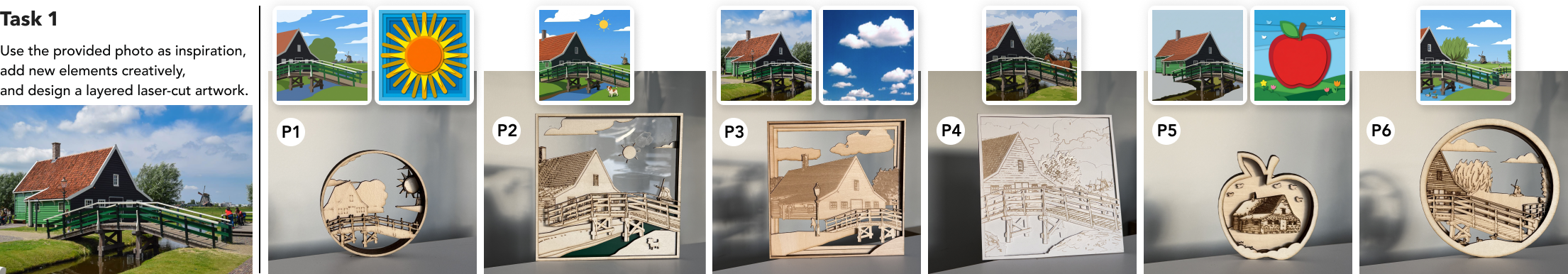}
  \caption{Task~1 outcomes. Top row: each participant's source image with added element(s). Bottom row: the fabricated artifact. All six participants created a layered souvenir from the same riverside village photograph.}
  \label{fig:study-outcomes}
  \Description{Top row: six panels showing each participant's source image with added elements. Bottom row: six corresponding fabricated layered laser-cut artifacts in various materials and frame shapes.}
\end{figure*}

\section{Practitioner Study}
\label{sec:evaluation}

Having validated \chal{C1--C3} (sourcing, preparation, spatial awareness) with novices and identified \chal{C4} (fabrication constraints) as unaddressed in Round~2 (Sec.~\ref{sec:formative}), we evaluated the redesigned FabDreamer with six fabrication practitioners spanning diverse domains and laser-cut expertise levels. The study addressed two questions: \textbf{RQ1}, whether FabDreamer effectively supports layered laser-cut art creation across expertise levels and resolves \chal{C1--C4}; and \textbf{RQ2}, how the three-stage workflow transfers to other fabrication domains and where domain-specific knowledge must be supplied.

\subsection{Participants} 

We recruited six fabrication practitioners (4~female, 2~male; ages 25--39) from the university community and local maker networks, selected for diversity in both fabrication domain and laser-cut expertise (Table~\ref{tab:participants}). All participants used AI tools in other aspects of their work, such as text generation or image creation, but none had integrated AI into their fabrication practice, citing a lack of tools that fit their fabrication workflows. Each participant was asked to bring 2--3 images from their own fabrication domain for our study.



\begin{table*}[t]
  \caption{Study participants. Yr = years of domain experience. LC = laser-cut experience (0: none, 1: observed, 2: occasional, 3: regular).}
  \label{tab:participants}
  \small\setlength{\tabcolsep}{3pt}
  \begin{tabular*}{\textwidth}{@{\extracolsep{\fill}}clrlc@{}}
    \toprule
    \textbf{ID} & \textbf{Role / Domain} & \textbf{Yr} & \textbf{Layered Artifact Practice} & \textbf{LC} \\
    \midrule
    P1 & Fabrication researcher & 10 & Wax-paper actuator prototypes & 2 \\
    P2 & Architecture instructor & 11 & Stacked contour landscape models & 2 \\
    P3 & Illustrator / sticker products & 4 & Layered scene stickers for DIY decor & 1 \\
    P4 & Makerspace manager / toy design & 13 & Pop-up greeting cards, multi-material & 3 \\
    P5 & Accessible tech researcher & 2 & Tactile Braille cards (fabric + paper) & 2 \\
    P6 & Fabric artist / children's books & 6 & Handmade layered fabric books & 0 \\
    \bottomrule
  \end{tabular*}
\end{table*}

\subsection{Study Protocol}

The study used a two-task within-participant design. In \textit{Task~1} (40--50 min), all participants created a layered laser-cut souvenir from the same riverside village photograph, each required to add at least one element not in the original image. In \textit{Task~2} (40--50 min), each participant chose one of the images they had brought from their own practice and used FabDreamer to work toward a domain-relevant artifact.
\add{The fixed Task~1 provides a controlled baseline; the contrast between each participant's behavior across the two tasks isolates domain transfer from system unfamiliarity and individual working style, and constitutes the central analytical move for RQ2.}

Each session lasted 90--120 minutes, conducted remotely via Zoom with participants on their own laptops. All sessions were screen- and audio-recorded. Before each task, a brief interview (10--15 min) established the participant's baseline workflow. After each task, a semi-structured interview (10--15 min) collected system impressions, \chal{C1--C4} verification (Task~1), and domain-transfer reflections (Task~2). Participants thought aloud throughout. \add{Interaction logs captured all AI invocations, canvas edits, and stage transitions, complementing the recordings and interviews.} Full interview protocols are provided in Appendix~\ref{appendix:study}.

\subsubsection{Fabrication and Follow-Up} 

In Stage~3, participants finalized material choices and physical dimensions; the research team then fabricated all 12 artifacts (6~participants $\times$ 2~tasks) on an xTool P2 CO$_2$ laser cutter at participant-specified scale, using manufacturer presets for the chosen material (3~mm basswood, 3~mm acrylic, poster paper, or foamcore). All artifacts were cut and assembled without structural failures. Participants received photographs and a short video of their fabricated artifacts and assessed, in a 10--15 minute follow-up interview, whether the physical result matched their design intent.

\subsubsection{Analysis.} Three researchers independently coded transcripts and think-aloud recordings following a reflexive thematic analysis approach~\cite{BraunClarke2006}, reconciling through discussion. For RQ1, we compared Task~1 behavior across expertise levels; for RQ2, we contrasted each participant's behavior across the two tasks to identify structural transfer versus domain-specific friction.

%% file: Sections/7-findings.tex
\subsection{Results}
\label{sec:findings}

\begin{figure*}[t]
  \centering
  \includegraphics[width=\textwidth]{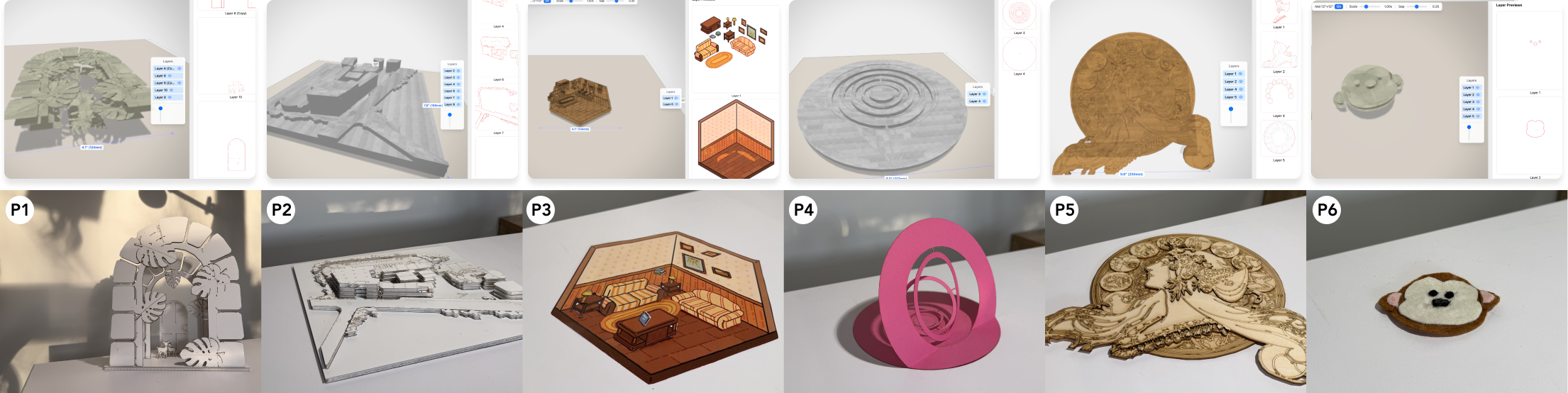} 
  \caption{Task~2 outcomes. Top row: each participant's design in FabDreamer's 3D preview. Bottom row: the fabricated artifact. Participants applied FabDreamer to six different fabrication domains.}
  \label{fig:task2-outcomes}
  \Description{Top row: six panels showing each participant's Task 2 design in FabDreamer's 3D preview. Bottom row: six corresponding fabricated artifacts spanning foamcore, paper, basswood, and fabric.}
\end{figure*}

All participants successfully completed both tasks and exported fabrication-ready files. All 12 artifacts were subsequently fabricated by the research team without structural failures (Figures~\ref{fig:study-outcomes}, \ref{fig:task2-outcomes}).

\subsubsection{\textbf{Task~1: Challenge Resolution Across Expertise Levels}}

All six participants completed Task~1 within 40--55 minutes regardless of laser-cut experience. \add{Despite working from the same photograph, designs diverged substantially in frame shape, material, scale, and added elements (Figure~\ref{fig:study-outcomes}); this divergence from identical input indicates a genuine creative design space rather than convergence toward a single solution.}

All four challenges were addressed, though with different resolution profiles. \chal{C1}~(sourcing) and \chal{C2}~(preparation) resolved uniformly: AI generation and the \textit{Grab} pathway removed element sourcing friction, with P1 noting the tool ``significantly lowers the cost of finding and preparing materials''; automated segmentation with on-demand repair addressed preparation bottlenecks, with P3 and P4 particularly valuing occlusion completion for reconstructing hidden content. \chal{C3}~(spatial awareness) also resolved uniformly but with expertise-dependent usage: all six participants used the 3D preview actively, though less experienced participants (P5, P6) relied on it for basic spatial orientation while more experienced ones (P2, P4) used it for precise alignment and material contrast decisions across layers.

\chal{C4}~(fabrication constraints) produced the most differentiated responses and most directly answers RQ1. Its resolution was universal but asymmetric across expertise levels. Participants with less laser-cut experience found agent warnings revelatory. P5's compact design triggered a thin-feature warning she had not anticipated; she scaled up the entire composition until it cleared. 
P2's sun was flagged as floating; rather than simply bridging it, he returned to Stage~2 and added a clear acrylic basement layer so the sun could appear suspended while remaining physically supported, turning a constraint warning into a material design decision. 
Participants with stronger fabrication expertise engaged warnings selectively. P4 rejected the auto-fix for her floating horse and drew her own bridge, controlling the attachment point for both structural and visual reasons. She also identified a constraint the agent had not flagged: auto-generated bridges may be too narrow to survive assembly handling even when they pass geometric checks, a gap between detected geometry and material survivability that expert users filled from their own knowledge.

\add{All six found FabDreamer more convenient and fluid than their self-described workflows, crediting the integrated pipeline for eliminating software-switching overhead (Sec.~\ref{sec:discussion}). Stage preferences were heterogeneous, consistent with the intent of varying AI initiative: P1 and P2 most valued Stage~1's generation pathway, P4 most valued Stage~3's prepare-and-preview workflow, and P3, P5, and P6 experienced the three stages as a coherent whole.}
Stage-level friction was concentrated in Stage~1 (Decompose) and modulated by familiarity with layered thinking: P2 and P4 confirmed AI proposals with minimal modification, while P5 and P6 \add{revised more iteratively} before proceeding. P4 summarized: ``The decompose part needs a little bit of scaffolding as a first-time user, but from create to prepare it's pretty straightforward.'' One Stage~3 (Prepare) gap emerged from post-task interviews: P1 observed that material choices made during design (thickness, burn thresholds) are not propagated into fabrication software settings, pointing toward a natural next step in closing the image-to-physical loop (Sec.~\ref{sec:limitations}).

\subsubsection{\textbf{Task~2: Workflow Transfer Across Domains}}

All six participants applied FabDreamer to their own domains and produced fabricable, domain-relevant designs (Figure~\ref{fig:task2-outcomes}, Table~\ref{tab:task2-workflow}). \add{Designs diverged substantially from Task~1: a 4-layer foamcore actuator prototype (P1), a 5-layer architectural site model from a plan drawing (P2), a 2.5D room-design sticker set (P3), a spiral pop-up greeting card (P4), a Mucha painting decomposed on 0.16\,mm basswood for tactile reading (P5), and a cartoon monkey in fur-color layers for fabric cutting (P6).}

\add{Two of these applications illustrate that the approach extends beyond decorative artifacts. P2, an architecture instructor, saw the integrated flow as removing the file-handoff overhead that dominates studio work, freeing students for design itself; for site models, working from images rather than meshes lowered the entry cost: ``3D models for slicing are very hard to find; comparatively, images are more convenient.'' P5, an accessibility researcher, positioned layered decomposition as a low-cost path to tactile graphics for blind and low-vision readers, a need text-only Braille tooling does not cover: ``we could make things that let them feel the brushstrokes... if there's a mountain, you actually make a raised mountain in the canvas.''}

Stages~1 (Decompose) and~2 (Create) transferred across all six domains with minimal friction, directly answering the first half of RQ2. The core workflow remained stable; what varied was how participants appropriated generic geometric operations for domain-specific purposes. P1 used the gap controller (for previewing layer spacing) to visualize how layers would look glued vertically onto a horizontal plane, repurposing a stacking tool for a non-stacking assembly. P2 duplicated building elements across multiple layers to express different building heights, reinterpreting layer depth as architectural elevation. P4 treated bridges not as constraint fixes but as structural joints for a fold-up pop-up card. P3 used FabDreamer for decomposition and repair, then exported PNGs into her existing Procreate-to-print pipeline, integrating the tool as an upstream stage rather than a complete replacement.


AI initiative comfort followed the Task~1 pattern across domains: all participants welcomed the Stage~1 layer proposal as a starting point, with the same split between minimal modification (P2, P4) and \add{iterative revision} (P3, P5, P6), and the Stage~2 on-demand model was accepted universally; P3 described the balance as natural, allowing her to focus on compositional decisions as she would in her own illustration software.

\begin{table}[t]
  \caption{Task~2 workflow adaptations by participants.}
  \label{tab:task2-workflow}
  \small
  \begin{tabular*}{\columnwidth}{@{}c@{\extracolsep{\fill}}p{7.2cm}@{}}
    \toprule
    \textbf{ID} & \textbf{Key Workflow Adaptation} \\
    \midrule
    P1 & Described full design in natural language; AI generated geometry directly. Used gap controller to preview vertical depth for layers glued onto a horizontal plane. \\
    \addlinespace
    P2 & Assigned site features to layers; duplicated building layers to simulate different heights. Toggled between Stage~2 and Stage~3 to iterate on layer count and material. \\
    \addlinespace
    P3 & Exported PNGs for all decomposed elements into existing print pipeline. Stage~2 Repair reconstructed occluded objects. SVG output eliminated manual bleed step. \\
    \addlinespace
    P4 & Created spiral shape, added bridges between concentric circles as structural joints. When folded, rings stand in 3D. Treated Prepare as part of Create. \\
    \addlinespace
    P5 & Decomposed complex organic artwork into semantically meaningful layers (jewelry, figure, zodiac, background). Fine engraving preserves brushstroke texture for tactile reading. \\
    \addlinespace
    P6 & Generated cartoon character, decomposed into fur-color and feature layers (muzzle, eyes, panels) for cutting from different colored cloth. \\
    \bottomrule
  \end{tabular*}
\end{table}

Stage~3 (Prepare) surfaced domain-specific constraints not encoded in the fabrication agent, directly answering the second half of RQ2. Four categories emerged from Task~2 interviews. \textit{Material parameters}: P1 and P4 both noted that cutting power and speed vary by material and machine, and that propagating material choices from design into machine settings would eliminate manual re-entry, a step the current export leaves to the user. \textit{Assembly and joinery}: P4 identified that how layers physically connect after cutting, whether by glue, notch, or press-fit, requires joinery knowledge the geometric agent cannot supply, referencing finger-joint and T-slot conventions the system does not represent. \textit{Tactile and spatial requirements}: P5's sensor assembly required a three-layer structure in which layer order was governed by circuit topology rather than visual depth, a constraint invisible to any geometry-only analysis. \textit{Print specifications}: P3 noted that her commercial print pipeline requires CMYK conversion and bleed width; the SVG export eliminated her manual bleed step but does not encode the color-profile metadata that print factories require.

\add{In follow-up interviews, all participants confirmed the physical output matched their design intent; the remaining assembly steps (P1's horizontal support plane, P4's fold pattern) drew on their own domain expertise.}

%% file: Sections/8-discussion.tex
\section{Discussion}
\label{sec:discussion}

\add{Though FabDreamer is instantiated in one domain, the cross-domain study lets us test, rather than assert, what generalizes (Sec.~\ref{sec:findings}). We draw three lessons for AI-assisted fabrication tools: carry fabrication constraints as a continuous thread rather than a terminal check; integrate the operations practitioners currently span across tools; and provide domain-neutral geometric operations that practitioners reinterpret.}

\subsection{Physical Awareness Creates Creative Opportunities}

\add{Making physical constraints visible during design supported creative invention beyond error prevention. Screen-based design tools assume that what looks right on screen is the end goal; when the target is a physical artifact, that assumption breaks down. FabDreamer's fabrication agent surfaced physical consequences while design decisions were still open, and participants responded by designing with the constraints: as seen with P2 and P5 in Task~1 (\chal{C4}), warnings became starting points for material and compositional choices. Here the constraint check acts as productive friction: an interruption that redirects rather than merely delays~\cite{chen2024exploring}.}

P4's engagement in Task~2 went further still. She recognized bridges not as constraint fixes but as structural elements for pop-up card design, connecting them to Love Pop's commercial products: ``Can I manipulate the bridge a little more?'' She iterated between Stage~2 and Stage~3, treating structural verification as part of creative composition rather than a final quality gate. 
This collapses the assumed boundary between ``create'' and ``prepare'', consistent with \citet{jacobs2018extending}, whose computational tools enhance practice without enforcing linear stage progression.




\subsection{End-to-End Integration Restructures Practice}

Practitioners valued the pipeline's integration more than the capability of any individual stage. Participants identified existing tools that handled individual stages better than FabDreamer (Rhino for vector editing, Photoshop for layer management, Cricut for material prompting), yet consistently endorsed the integrated pipeline. P2 explained: ``Switching between software is very troublesome; this kind of integrated flow is obviously better, and file handoff between tools is itself a common source of error,'' echoing findings on hybrid craft tools where workflow continuity outweighs per-stage capability~\cite{zoran2013hybrid}.

P3 went further: with FabDreamer, she ``probably wouldn't bother separating layers in detail while drawing; I'd just throw it in here to separate at the end.'' This comment concerns upstream behavior rather than decomposition quality: knowing that the pipeline handles decomposition changes how she would draw from the outset. Similarly, P2 observed that layer planning in Stage~1 ``seems not that important'' because Stage~2 allows reassignment. The tool does not merely automate steps in an existing workflow; it restructures how practitioners think about their entire pipeline.

\add{This restructuring is not reducible to efficiency gains: it appears precisely because physical feedback (3D preview, constraint detection, material simulation) lives in the same environment as image editing and layer composition. The design implication is that integration should be a first-order goal for image-to-physical tools, with its payoff measured upstream as well as downstream~\cite{campbell2025effort}. Their preference for the pipeline over stronger standalone stages mirrors the distinction in behavioral design between optimizing individual choices and shaping the infrastructure that conditions them~\cite{schmidt2022choice}.}

\subsection{Geometric Primitives Are Domain-Appropriable}

Practitioners appropriate generic geometric primitives for domain-specific purposes the system was never designed for, which means supporting multiple fabrication domains does not require encoding each domain's rules into the agent. P4 repurposed bridges as pop-up card joints, P1 repurposed the gap controller as a vertical-assembly preview, and P2 repurposed layer depth as architectural elevation, in each case supplying the domain interpretation that the system's geometry-only representation left open.

The extensibility model this suggests is to provide appropriable geometric operations and let practitioners supply the interpretation. This aligns with \citet{li2023beyond}'s call for vertical movement in creativity support tools: the ability to inspect and retool a tool's abstractions rather than being bound by them, applied here to physical fabrication.

\add{The constraint taxonomy in Sec.~\ref{sec:findings}, however, marks the boundary of this model: the four domain-specific constraint types practitioners identified cannot be detected from geometry alone. P5's trust concern (``If AI helped me, I would expect it to definitely work'') is instructive: an agent that flags floating elements with the same confidence as it estimates joinery clearance misleads practitioners exactly when they need calibrated judgment. The boundary between geometry-detectable and domain-supplied constraints should be a visible property of the agent interface~\cite{li2023beyond}, as in other review tools where the agent surfaces candidate issues for expert judgment rather than resolving them~\cite{chen2026privacymotiv}.}

\section{Limitations and Future Work}
\label{sec:limitations}

\add{The decomposition pipeline relies on SAM2 with generative extraction as fallback: SAM2 can confidently segment the wrong element when bounding boxes overlap, occlusion completion produces plausible but not faithful output, and because each layer's inpainting feeds the next, errors can compound; the modular architecture allows swapping in improved models. Our study included six participants in single sessions: the domain diversity provided breadth for cross-domain patterns, but longitudinal use would reveal whether the adaptations we observed persist beyond initial exposure.}

\add{Stage~1's text-based proposal could be complemented by interaction on the segmentation itself, such as direct click prompts augmenting the VLM's prompts to SAM2 or multiple candidate decompositions to choose among; a segment-first, group-later pathway would invert the order-then-segment architecture that enables occlusion inpainting (Sec.~\ref{subsec:decomposition}). Offering several candidates rather than one would also change what users explore before committing, an effect documented for pre-decision exploration designs~\cite{chen2026framing}.}

\add{Our account of the varying-initiative design is qualitative (Sec.~\ref{sec:findings}), and the right initiative setting proved domain-dependent: P1 weighted workflow efficiency, P3 creative control. A standardized per-stage scale would average away exactly these differences, so a systematic initiative-allocation study is a natural next step.}

\add{Within the geometry-detectable class the agent covers, further checks such as kerf-aware spacing and press-fit clearances can join the agent's detection loop (Sec.~\ref{subsec:fabagent}). FabDreamer's output is a fabrication-ready SVG; machine-specific parameters remain with fabrication software presets, and propagating material choices into machine settings would close this remaining gap in the image-to-physical workflow. Both pilot experts and participants (P4) highlighted FabDreamer's potential for scaffolding students' understanding of the digital-to-physical transition.}


%% file: Sections/9-conclusion.tex

\section{Conclusion}

Developing FabDreamer and observing six practitioners apply it within their own fabrication domains yields three conclusions that extend beyond the system itself.
\add{First, AI initiative should be calibrated to the recoverability of error: participants accepted an AI that leads when errors are recoverable, and expected to lead themselves when errors consume material. This calibration, more than any single capability, sustained the collaboration, and it offers designers of fabrication tools a concrete criterion for deciding where automation belongs.
Second, visible physical constraints open creative opportunities beyond error prevention: when structural warnings carried physical explanations, practitioners treated them as design material, inventing support layers and repurposing bridges as structural joints. Constraint feedback should therefore be designed as a creative surface woven through composition rather than as a terminal check before export.
Third, domain-neutral geometric operations scale across domains in a way that per-domain rules cannot: practitioners supplied the domain meaning themselves, reading layer depth as architectural elevation and gap spacing as assembly preview. Designers of fabrication agents should invest in appropriable primitives and make visible the boundary between what geometry can verify and what only the maker knows.
The image-to-physical gap will continue to narrow as models improve. How initiative, constraints, and domain knowledge are shared between maker and machine is the more durable design question, and these three conclusions constitute our answer.}

%% file: Sections/appendix.tex
\section{Round 1: Formative Study Details}
\label{appendix:formative}

Following methods from prior analyses of creative practices~\cite{liao2022realitytalk, li2023stargazer}, we analyzed 25 YouTube tutorials (2019--2024) on multilayered laser-cut design workflows, sourced using keywords including ``multilayered laser cut workflow'' and ``layered laser cut tutorial.'' We excluded content focused solely on machine operation, single-layer cutting, or assembly without design documentation. We additionally conducted a 60-minute semi-structured interview with a professional publisher with over 10 years of experience in digital fabrication and layered laser-cut production. Guided by Albaugh et al.'s material practice framework~\cite{albaugh2024hybrid}, three researchers independently coded the video content and interview transcript across three dimensions: physical characteristics and operations, abstractions and workflows, and aesthetics and forms. A reflexive thematic analysis of the coded data produced the five-phase workflow (ideating, preparing, arranging, processing, finalizing) and the four challenges described in Section~\ref{sec:formative}. Scenic and landscape designs were the most common category (60\% of tutorials), and 76\% of creators relied on existing vector libraries rather than personal images for source material.

\section{Round 2: Early Prototype Study Details}
\label{appendix:pilot}

\subsection{Early Prototype}

The prototype implemented a three-module pipeline for converting images into laser-ready layered designs. An \textit{image generation module} produced images from text prompts using Flux.1~\cite{BlackForestLabs2024}, with prompt engineering to encourage clearly segmentable entities. A \textit{layer decomposition module} used a vision-language model (Qwen2-VL-72B~\cite{Qwen2-VL}) to analyze depth relationships and assign entities to up to five layers, then applied Grounding DINO~\cite{ren2024grounding} for entity detection, SAM2~\cite{ravi2024sam2} for segmentation masks, and Stable Diffusion 3 inpainting~\cite{esser2024scalingrectifiedflowtransformers} to fill occluded regions. A \textit{fabrication preparation module} applied style transformations (mosaic or line art) and converted layers to SVG files with color-coded cut and engrave paths.

The web interface provided five panels: image generation and upload, entity segmentation display, basic editing controls (scale, rotate, position), drag-and-drop layer management, and a real-time 3D stacked preview. Users could generate or upload images, view auto-decomposed layers, rearrange entities across layers, apply one of two preset styles, and export SVG files for laser cutting. The system did not support error recovery during segmentation, offered limited creative editing, and performed no fabrication constraint checking.

\subsection{Round 2 User Study Procedure}

We conducted a two-phase pilot study to evaluate the prototype with both novice users and domain experts.

In \textbf{Phase 1}, we recruited 10 participants (5 female, 5 male; mean age = 23.9, SD = 3.1) with no prior experience in layered laser-cut design. Participants included graphic designers (3), digital artists (2), and hobbyist makers (5). Each completed a 60-minute remote session: a brief introduction to layered laser-cut art (10 minutes), an open-ended design task on a self-selected topic with think-aloud protocol (35--40 minutes), and a post-task survey and semi-structured interview (10 minutes). We documented screen interactions, system logs, and verbal explanations. All designs were fabricated on a Universal Laser System ULTRA X6000, cutting 5.5-inch square basswood panels at 1/8-inch thickness, and participants reviewed their physical results in a follow-up session.

In \textbf{Phase 2}, three domain experts---an educator with over 10 years teaching digital fabrication, a makerspace director specializing in creative fabrication, and an active maker with five years of layered art experience---each reviewed the fabricated pieces and recorded design processes in a 60-minute session, providing structured feedback on design quality, technical viability, and system potential.

\begin{figure*}[t]
  \centering
  \includegraphics[width=\textwidth]{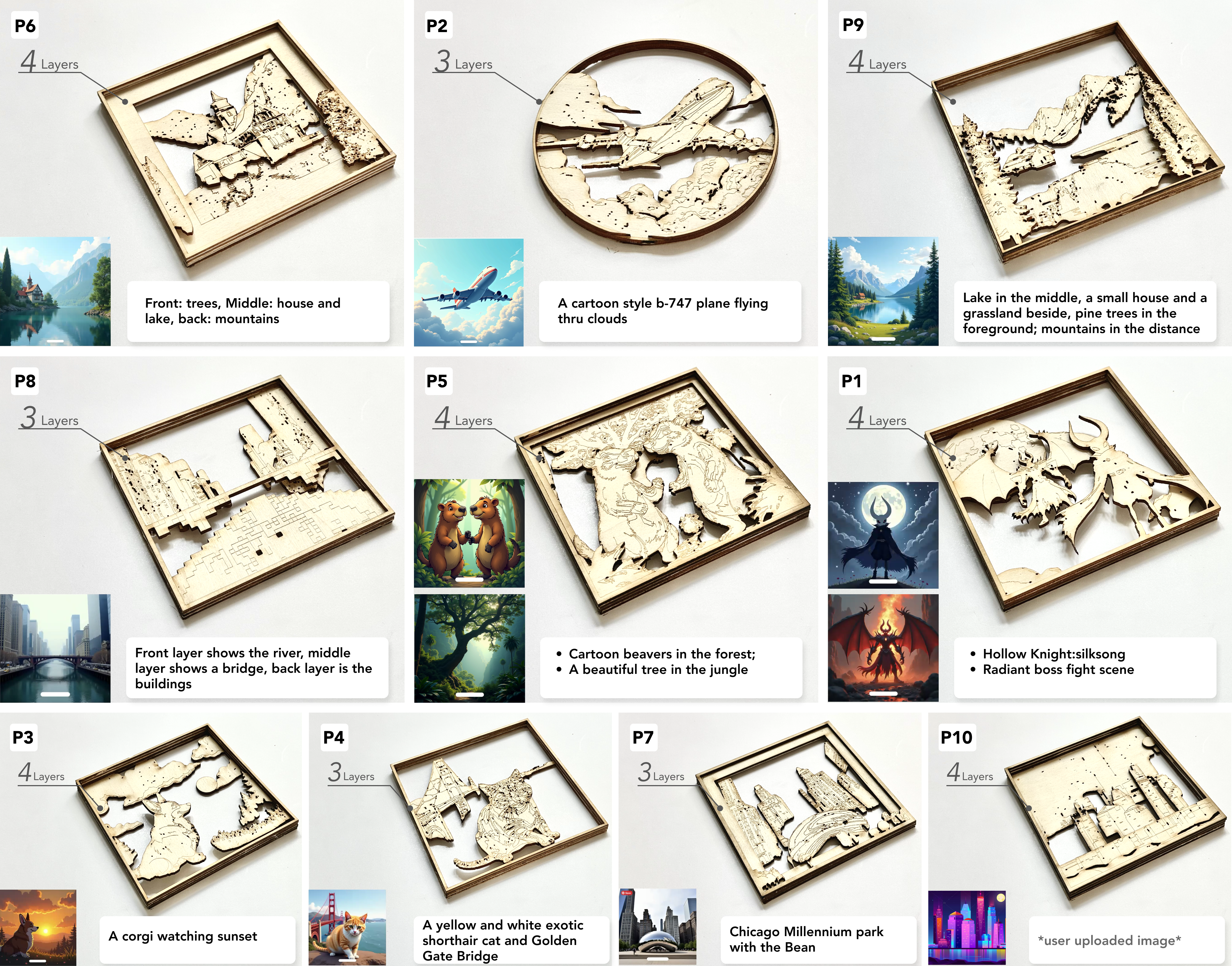}
  \caption{Round 2 user study participant creations. All 10 novice participants successfully created 3--5 layer designs spanning diverse subjects---landscapes, character scenes, architectural compositions---and all designs survived laser cutting and physical assembly. Each cell shows the text prompt, generated image, and final fabricated artifact.}
  \label{fig:pilot-results}
  \Description{A grid of pilot study participant outcomes. Each cell shows a text prompt, the AI-generated image, and the final laser-cut layered artifact. Designs include landscapes, cityscapes, character scenes, and architectural compositions, all successfully fabricated as 3--5 layer stacked wood pieces.}
\end{figure*}

\subsection{Round 2 Results}
\label{appendix:pilot-results}

All 10 participants successfully created multi-layered designs ranging from 3--5 layers, completing their designs in an average of 25 minutes (Figure~\ref{fig:pilot-results}). Subjects spanned character and animal scenes, natural landscapes, and architectural compositions---all of which survived laser cutting and physical assembly. Survey ratings were positive overall, with the real-time 3D preview feature receiving the highest satisfaction (mean = 4.9/5), followed by layer control feature (4.0/5) and entity editing feature (3.8/5). These outcomes validated that the core pipeline---from AI-generated image through layer decomposition to physical fabrication---was viable. The gaps identified from these findings are described in Sec.~\ref{sec:formative}.

\section{Round 3: Practitioner Study Details}
\label{appendix:study}

\subsection{Recruitment}

In Table~\ref{tab:participants}, \textbf{Yr} denotes years of experience in the participant's primary fabrication domain. \textbf{Layered Artifact Practice} describes each participant's layered artifact work in their own domain, from which they brought 2--3 images for Task~2. \textbf{LC} (laser-cut experience) is rated on a four-point scale: 0 = no experience; 1 = observed others but never operated independently; 2 = occasional use, a few times per year depending on project needs; 3 = regular use, several times per month.

Candidates were screened using an online eligibility form
covering primary fabrication domain, years of practice, types
of layered artifacts produced, and current software tools. We
required at least two years of active fabrication practice and
at least one layered physical artifact produced in the
candidate's domain. We excluded candidates whose primary
practice was laser cutting specifically, to avoid a ceiling
effect in Task~1 that would reduce the behavioral contrast
between tasks. All participants used AI tools in other
contexts (text generation, image generation, or design
assistance) but had not integrated AI into their fabrication
workflows, confirming they were neither resistant to AI in
principle nor already habituated to AI-assisted fabrication.

\subsection{Background Survey}

Administered online before the session to preserve session
time for qualitative data collection. Four sections:

\textbf{Practice.} Role, years of fabrication experience,
primary domain, materials, and machines used.

\textbf{Layered artifact experience.} Whether participants had
created layered physical artifacts, a description of a recent
example, and their typical image-to-physical workflow
including most time-consuming and most technically difficult
steps.

\textbf{Laser-cut experience.} Rated on a four-point scale:
0~=~no experience; 1~=~observed others but never operated
independently; 2~=~occasional use, a few times per year
depending on project needs; 3~=~regular use, several times
per month.

\textbf{AI familiarity.} Frequency of use, specific tools
used, self-assessed comfort level, and whether participants
had used any AI tool for image decomposition or segmentation.

\subsection{Round 3 Procedure and Interview Protocols}

\textbf{Opening Interview} (15 minutes).
Participants described a recent fabrication project from start
to finish before any system interaction, with probes for \chal{C1--C4}
if not raised spontaneously. For example, \textit{``Have
you ever had a design fail during fabrication because of
something not caught at the design stage?''} The vocabulary
that emerged served as a pre-tool baseline for RQ2 analysis.

\textbf{Pre-Task~1 Interview} (10 minutes).
Participants were shown the Task~1 image and asked to walk
through how they would approach it using their own tools,
including estimated time and hardest step. The interviewer
probed for \chal{C1--C4} if not mentioned. Estimated times and
identified hardest steps were recorded as anchors for
post-task challenge verification.

\textbf{Post-Task~1 Interview} (10--15 minutes).
System impressions were anchored to each participant's
pre-task estimates: \textit{``Earlier you estimated this would
take [X] and that [step] would be hardest --- how did your
experience compare?''} \chal{C1--C4} challenge verification followed,
with the \chal{C4} probe: \textit{``Did the fabrication agent catch
things you would have caught yourself, or things you would not
have noticed? Did it miss anything you knew would be a
problem?''} Feature-level feedback covered the layer proposal
in Stage~1, the canvas and 3D preview in Stage~2, and the
fabrication agent and SVG export in Stage~3. Feature feedback
was collected here rather than after Task~2 because
participants had just engaged with every feature in a
controlled, fixed-image context.

\textbf{Pre-Task~2 Interview} (10 minutes).
Same \chal{C1--C4} probe structure as the pre-Task~1 interview but
focused on the participant's primary domain. Two activation
questions surfaced initial transfer mental models before
system exposure: \textit{``How would you describe FabDreamer
to a colleague in your field?''} and \textit{``If you were
going to use FabDreamer on a project in your domain, what
kind of project would you try?''}

\textbf{Post-Task~2 Interview} (10--15 minutes).
Transfer impressions: which stages transferred directly, which
required adaptation, and which felt absent or mismatched.
Domain-specific constraint probe: \textit{``Were there
fabrication constraints that matter in your domain that
FabDreamer had no concept of? How should a tool communicate
those to someone without your expertise?''} Task contrast:
\textit{``Did you use the tool differently in Task~2 than in
Task~1? What constraints were you thinking about that the tool
did not ask you about?''}

\textbf{Follow-Up Interview} (10--15 minutes).
Conducted after participants received photographs and a short
video of their fabricated artifacts. Physical output
assessment: whether the result matched design intent, any
structural surprises, and what would need to be different for
the artifact to be usable in their actual practice.
Revised overall assessment: \textit{``Now that you have seen
the physical results, has your assessment of FabDreamer
changed?''}

\section{AI Components and Prompt Design}
\label{appendix:llm}


FabDreamer uses 19 AI components plus SAM2 (Table~\ref{tab:ai-components}). All Gemini calls use the Google Gemini API via the \texttt{google-genai} SDK. The system uses three model tiers: Gemini~3.1 Pro for image analysis and spatial reasoning, Gemini~3.1 Flash Image for generation and inpainting, and Gemini~2.5 Flash for the fabrication agent. SAM2 Large is self-hosted on a Mac Mini (M4) for pixel-level segmentation. Each component includes an automatic fallback to an alternative model on API errors.

\begin{table}[t]
\centering
\caption{AI component summary. Components 1--17 operate in Stage~1; component~18 in Stage~2; component~19 in Stage~3.}
\label{tab:ai-components}
\scriptsize\setlength{\tabcolsep}{3pt}
\begin{tabular*}{\columnwidth}{@{\extracolsep{\fill}}rlll@{}}
\toprule
\# & Component & Model & Output \\
\midrule
1 & Layer analysis & \texttt{gemini-3.1-pro} & JSON \\
2 & Batch detection & \texttt{gemini-3.1-pro} & JSON \\
3 & Grab detection & \texttt{gemini-3.1-pro} & JSON \\
4 & Sub-segment detection & \texttt{gemini-3.1-pro} & JSON \\
5 & Humanize names & \texttt{gemini-2.5-flash} & JSON \\
6 & Layer modification & \texttt{gemini-3.1-pro} & JSON \\
7 & Rough mask (diff-mask) & \texttt{gemini-2.5-flash-image} & Image \\
8 & Native mask (fallback) & \texttt{gemini-2.5-flash-image} & Image \\
9 & Image structure analysis & \texttt{gemini-3.1-pro} & Text \\
10 & Per-layer inpaint analysis & \texttt{gemini-3.1-pro} & Text \\
11 & Background inpainting & \texttt{gemini-3.1-flash-image} & Image \\
12 & Occlusion check (simple) & \texttt{gemini-3.1-pro} & JSON \\
13 & Occlusion check (detailed) & \texttt{gemini-3.1-pro} & JSON \\
14 & Occlusion completion & \texttt{gemini-3.1-flash-image} & Image \\
15 & Image generation & \texttt{gemini-3.1-flash-image} & Image \\
16 & Image update & \texttt{gemini-3.1-flash-image} & Image \\
17 & Image stylization & \texttt{gemini-3.1-flash-image} & Image \\
18 & Segment repair & \texttt{gemini-3.1-flash-image} & Image \\
19 & Fabrication agent & \texttt{gemini-2.5-flash} & Text + tools \\
--- & SAM2 Large & \texttt{sam2.1\_hiera\_large} & Mask \\
\bottomrule
\end{tabular*}
\end{table}

\begin{figure*}[t]
  \centering
  \includegraphics[width=\textwidth]{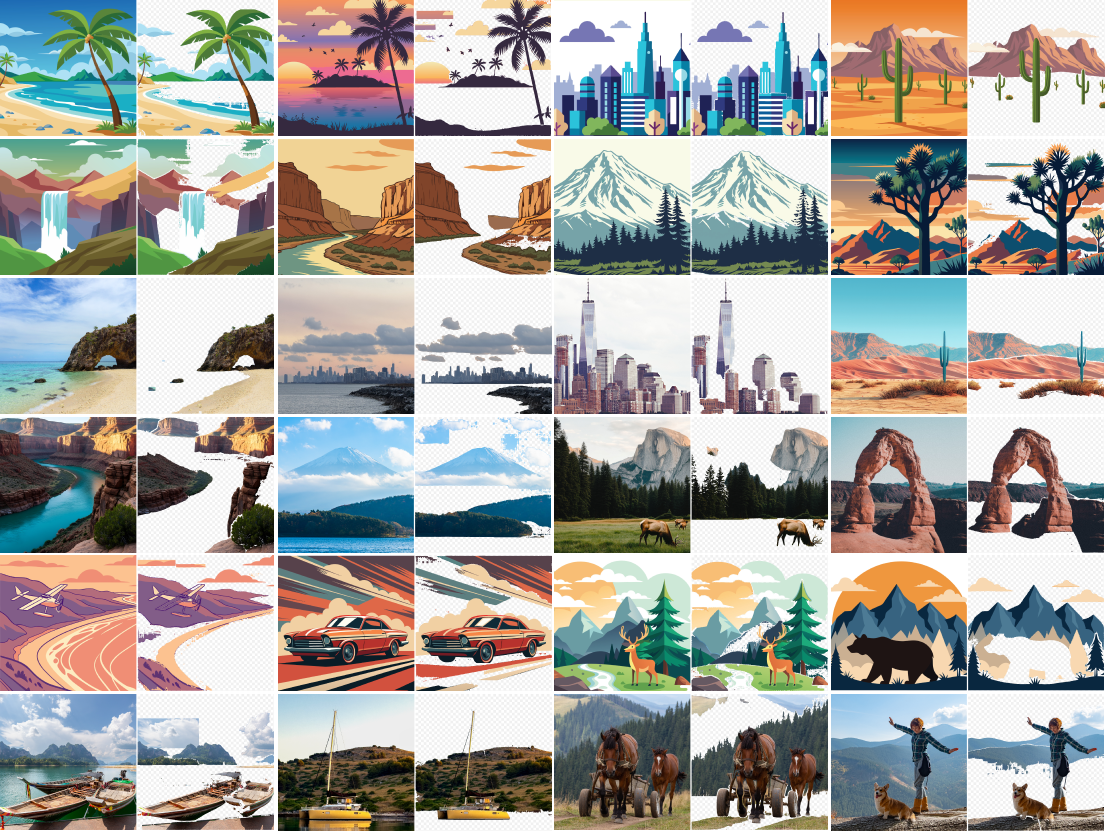}
  \caption{Pipeline test results for 24~images. Each pair shows the input image (left) and the one-shot cascade result with extracted elements stacked (right). Most noisy elements can be recovered via the repair tool in Stage~2. Low-contrast elements such as distant mountains and clouds against light skies are the most challenging cases for extraction.}
  \label{fig:pipeline-eval}
  \Description{Grid of 24 image pairs. Each pair shows the input image on the left and stacked extracted elements on the right. Images span real photographs and AI-generated illustrations of landscapes (scenery, city, desert, beach) and objects with background (animals, transportation).}
\end{figure*}

Two prompt design patterns are worth noting. First, because the pipeline removes elements front-to-back and inpaints between layers, earlier elements may be gone by the time deeper layers are analyzed. All element descriptions use absolute spatial references (``bottom-center of the image'') rather than relative ones (``below the duck''), and each inpainting prompt lists previously removed elements to prevent the generator from recreating them. Second, background reconstruction uses a two-level strategy: a one-time style analysis captures the input's art style and texture language (e.g., ``quilted fabric art, crosshatch linen background''), and per-layer fill instructions reference this description to maintain visual consistency (e.g., ``continue red-navy plaid pattern at 45-degree diagonal''). Occlusion completion and segment repair draw the target element on a solid magenta background; chroma-key removal produces a clean transparent PNG whose alpha channel serves as the segmentation mask. We also explored an alternative all-Gemini pipeline (${\sim}$1,500 lines) that used the image generation model for both detection and mask generation in a single multi-turn session, eliminating SAM2. It was abandoned after producing lower mask quality than the hybrid approach for pixel-precise segmentation.

\section{Pipeline Evaluation Details}
\label{appendix:pipeline-eval}

\add{The 24~test images were} downloaded from Vecteezy (free license) and manually cropped to $1024 \times 1024$ pixels, spanning six subcategories: scenery (6), city (2), desert (4), beach (4), animals (4), and transportation (4)\add{; extraction-route shares across the 107~layers and 238~elements are reported in Table~\ref{tab:pipeline-eval}. Generative extraction handled elements with structural openings (e.g., tree canopies, rock arches); the bounding-box fallback cases were primarily low-contrast clouds against light skies.} Illustration and vector-style images produced the cleanest extractions due to sharp object boundaries. For real photographs, animals and human figures segmented well, consistent with SAM2's training distribution, while diffuse natural boundaries (distant mountains, fog gradients, low-contrast terrain) were the most challenging (Figure~\ref{fig:pipeline-eval}).

%% file: reference.bib
@String{Computing = "Computing" }

@String{Computer = "{IEEE} Computer" }

@String{Springer = "Springer-Verlag" }

@inproceedings{umapathi2015laserstacker,
  title={LaserStacker: Fabricating 3D objects by laser cutting and welding},
  author={Umapathi, Udayan and Chen, Hsiang-Ting and Mueller, Stefanie and Wall, Ludwig and Seufert, Anna and Baudisch, Patrick},
  booktitle={Proceedings of the 28th Annual ACM Symposium on User Interface Software \& Technology},
  pages={575--582},
  year={2015},
  doi = {10.1145/2807442.2807512}
}

@inproceedings{mueller2013laserorigami,
  title={LaserOrigami: laser-cutting 3D objects},
  author={Mueller, Stefanie and Kruck, Bastian and Baudisch, Patrick},
  booktitle={Proceedings of the SIGCHI Conference on Human Factors in Computing Systems},
  pages={2585--2592},
  year={2013},
  doi = {10.1145/2470654.2481358}
}

@inproceedings{faruqi2023style2fab,
  title={Style2Fab: Functionality-Aware Segmentation for Fabricating Personalized 3D Models with Generative AI},
  author={Faruqi, Faraz and Katary, Ahmed and Hasic, Tarik and Abdel-Rahman, Amira and Rahman, Nayeemur and Tejedor, Leandra and Leake, Mackenzie and Hofmann, Megan and Mueller, Stefanie},
  booktitle={Proceedings of the 36th Annual ACM Symposium on User Interface Software and Technology},
  pages={1--13},
  year={2023},
  doi = {10.1145/3586183.3606723}
}

@inproceedings{qian2024shape,
  title={SHAPE-IT: Exploring Text-to-Shape-Display for Generative Shape-Changing Behaviors with LLMs},
  author={Qian, Wanli and Gao, Chenfeng and Sathya, Anup and Suzuki, Ryo and Nakagaki, Ken},
  booktitle={Proceedings of the 37th Annual ACM Symposium on User Interface Software and Technology},
  pages={1--29},
  year={2024},
  doi = {10.1145/3654777.3676348}
}

@inproceedings{baudisch2019kyub,
  title={Kyub: A 3d editor for modeling sturdy laser-cut objects},
  author={Baudisch, Patrick and Silber, Arthur and Kommana, Yannis and Gruner, Milan and Wall, Ludwig and Reuss, Kevin and Heilman, Lukas and Kovacs, Robert and Rechlitz, Daniel and Roumen, Thijs},
  booktitle={Proceedings of the 2019 CHI Conference on Human Factors in Computing Systems},
  pages={1--12},
  year={2019},
  doi = {10.1145/3290605.3300796}
}

@misc{openai_dalle,
  author = {OpenAI},
  title = {{DALL·E: Creating Images from Text}},
  year = {2021},
  howpublished = {\url{https://openai.com/dall-e}},
  note = {Accessed: 2024-04-02}
}

@inproceedings{sethapakdi2021fabricaide,
  title={Fabricaide: Fabrication-aware design for 2d cutting machines},
  author={Sethapakdi, Ticha and Anderson, Daniel and Sy, Adrian Reginald Chua and Mueller, Stefanie},
  booktitle={Proceedings of the 2021 CHI Conference on Human Factors in Computing Systems},
  pages={1--12},
  year={2021},
  doi = {10.1145/3411764.3445345}
}

@inproceedings{park2022foolproofjoint,
  title={FoolProofJoint: Reducing Assembly Errors of Laser Cut 3D Models by Means of Custom Joint Patterns},
  author={Park, Keunwoo and Lempert, Conrad and Abdullah, Muhammad and Katakura, Shohei and Shigeyama, Jotaro and Roumen, Thijs and Baudisch, Patrick},
  booktitle={Proceedings of the 2022 CHI Conference on Human Factors in Computing Systems},
  pages={1--12},
  year={2022},
  doi = {10.1145/3491102.3501919}
}

@inproceedings{roumen2021autoassembler,
  title={autoAssembler: Automatic Reconstruction of Laser-Cut 3D Models},
  author={Roumen, Thijs and Lempert, Conrad and Apel, Ingo and Brendel, Erik and Brand, Markus and Seidel, Laurenz and Rambold, Lukas and Baudisch, Patrick},
  booktitle={The 34th Annual ACM Symposium on User Interface Software and Technology},
  pages={652--662},
  year={2021},
  doi = {10.1145/3472749.3474776}
}

@inproceedings{beyer2015platener,
  title={Platener: Low-fidelity fabrication of 3D objects by substituting 3D print with laser-cut plates},
  author={Beyer, Dustin and Gurevich, Serafima and Mueller, Stefanie and Chen, Hsiang-Ting and Baudisch, Patrick},
  booktitle={Proceedings of the 33rd Annual ACM Conference on Human Factors in Computing Systems},
  pages={1799--1806},
  year={2015},
  doi = {10.1145/2702123.2702225}
}

@inproceedings{nisser2021laserfactory,
  author = {Nisser, Martin and Liao, Christina and Chai, Yuchen and Adhikari, Aradhana and Hodges, Steve and Mueller, Stefanie},
  title = {{LaserFactory}: A Laser Cutter-based Electromechanical Assembly and Fabrication Platform to Make Functional Devices and Robots},
  booktitle = {Proceedings of the 2021 CHI Conference on Human Factors in Computing Systems},
  pages = {1--15},
  year = {2021},
  doi = {10.1145/3411764.3445692},
  publisher = {ACM}
}

@misc{Qwen2-VL,
  author = {Wang, Peng and Bai, Shuai and Tan, Sinan and Wang, Shijie and Fan, Zhihao and others},
  title = {{Qwen2-VL}: Enhancing Vision-Language Model's Perception of the World at Any Resolution},
  year = {2024},
  eprint = {2409.12191},
  archivePrefix = {arXiv},
  primaryClass = {cs.CV},
  url = {https://arxiv.org/abs/2409.12191}
}

@misc{ren2024grounding,
  title={Grounding DINO 1.5: Advance the "Edge" of Open-Set Object Detection}, 
  author={Tianhe Ren and Qing Jiang and Shilong Liu and Zhaoyang Zeng and Wenlong Liu and Han Gao and Hongjie Huang and Zhengyu Ma and Xiaoke Jiang and Yihao Chen and Yuda Xiong and Hao Zhang and Feng Li and Peijun Tang and Kent Yu and Lei Zhang},
  year={2024},
  eprint={2405.10300},
  archivePrefix={arXiv},
  primaryClass={cs.CV}
}

@article{ravi2024sam2,
  title={SAM 2: Segment Anything in Images and Videos},
  author={Ravi, Nikhila and Gabeur, Valentin and Hu, Yuan-Ting and Hu, Ronghang and Ryali, Chaitanya and Ma, Tengyu and Khedr, Haitham and R{\"a}dle, Roman and Rolland, Chloe and Gustafson, Laura and Mintun, Eric and Pan, Junting and Alwala, Kalyan Vasudev and Carion, Nicolas and Wu, Chao-Yuan and Girshick, Ross and Doll{\'a}r, Piotr and Feichtenhofer, Christoph},
  journal={arXiv preprint arXiv:2408.00714},
  url={https://arxiv.org/abs/2408.00714},
  year={2024}
}

@misc{esser2024scalingrectifiedflowtransformers,
  title={Scaling Rectified Flow Transformers for High-Resolution Image Synthesis}, 
  author={Patrick Esser and Sumith Kulal and Andreas Blattmann and Rahim Entezari and Jonas Müller and Harry Saini and Yam Levi and Dominik Lorenz and Axel Sauer and Frederic Boesel and Dustin Podell and Tim Dockhorn and Zion English and Kyle Lacey and Alex Goodwin and Yannik Marek and Robin Rombach},
  year={2024},
  eprint={2403.03206},
  archivePrefix={arXiv},
  primaryClass={cs.CV},
  url={https://arxiv.org/abs/2403.03206}, 
}

@misc{rombach2021highresolution,
  title={High-Resolution Image Synthesis with Latent Diffusion Models}, 
  author={Robin Rombach and Andreas Blattmann and Dominik Lorenz and Patrick Esser and Björn Ommer},
  year={2021},
  eprint={2112.10752},
  archivePrefix={arXiv},
  primaryClass={cs.CV}
}

@misc{midjourney,
  title = {Midjourney (V5.1)},
  author = {{Midjourney}},
  year = {2023},
  url = {https://www.midjourney.com/}
}

@misc{glowforge,
  author = {{Glowforge}},
  title = {Bring your ideas to life.},
  year = {2024},
  url = {https://glowforge.com/},
}

@misc{BlackForestLabs2024,
  author = {{Black Forest Labs}},
  title = {Announcing {Black Forest Labs}},
  year = {2024},
  howpublished = {\url{https://blackforestlabs.ai/announcing-black-forest-labs/}},
  note = {Accessed: 2024-08-02}
}

@inproceedings{liao2022realitytalk,
  title={Realitytalk: Real-time speech-driven augmented presentation for ar live storytelling},
  author={Liao, Jian and Karim, Adnan and Jadon, Shivesh Singh and Kazi, Rubaiat Habib and Suzuki, Ryo},
  booktitle={Proceedings of the 35th annual ACM symposium on user interface software and technology},
  pages={1--12},
  year={2022},
  doi = {10.1145/3526113.3545702}
}

@inproceedings{li2023stargazer,
  title={Stargazer: An interactive camera robot for capturing how-to videos based on subtle instructor cues},
  author={Li, Jiannan and Sousa, Maur{\'\i}cio and Mahadevan, Karthik and Wang, Bryan and Aoyagui, Paula Akemi and Yu, Nicole and Yang, Angela and Balakrishnan, Ravin and Tang, Anthony and Grossman, Tovi},
  booktitle={Proceedings of the 2023 CHI Conference on Human Factors in Computing Systems},
  pages={1--16},
  year={2023},
  doi = {10.1145/3544548.3580896}
}

@article{koo2016towards,
  title={Towards zero-waste furniture design},
  author={Koo, Bongjin and Hergel, Jean and Lefebvre, Sylvain and Mitra, Niloy J},
  journal={IEEE transactions on visualization and computer graphics},
  volume={23},
  number={12},
  pages={2627--2640},
  year={2016},
  publisher={IEEE},
  doi = {10.1109/TVCG.2016.2633519}
}

@inproceedings{leake2024scrapmap,
  title={ScrapMap: Interactive Color Layout for Scrap Quilting},
  author={Leake, Mackenzie and Daly, Ross},
  booktitle={Proceedings of the 37th Annual ACM Symposium on User Interface Software and Technology},
  pages={1--17},
  year={2024},
  doi = {10.1145/3654777.3676404}
}

@inproceedings{liu20233dall,
  title={3DALL-E: Integrating text-to-image AI in 3D design workflows},
  author={Liu, Vivian and Vermeulen, Jo and Fitzmaurice, George and Matejka, Justin},
  booktitle={Proceedings of the 2023 ACM designing interactive systems conference},
  pages={1955--1977},
  year={2023},
  doi = {10.1145/3563657.3596098}
}

@inproceedings{chen2024exploring,
  title={Exploring a Behavioral Model of “Positive Friction” in Human-AI Interaction},
  author={Chen, Zeya and Schmidt, Ruth},
  booktitle={International Conference on Human-Computer Interaction},
  pages={3--22},
  year={2024},
  organization={Springer},
  doi = {10.1007/978-3-031-61353-1_1}
}

@inproceedings{shen2024neural,
  title={Neural canvas: Supporting scenic design prototyping by integrating 3d sketching and generative AI},
  author={Shen, Yulin and Shen, Yifei and Cheng, Jiawen and Jiang, Chutian and Fan, Mingming and Wang, Zeyu},
  booktitle={Proceedings of the 2024 CHI Conference on Human Factors in Computing Systems},
  pages={1--18},
  year={2024},
  doi = {10.1145/3613904.3642096}
}

@inproceedings{albaugh2024hybrid,
  author = {Albaugh, Lea and Gonzalez, Jesse T and Hudson, Scott E},
  title = {Tensions and Resolutions in Hybrid Basketry: Joining 3D Printing and Handweaving},
  booktitle = {Proceedings of the Eighteenth International Conference on Tangible, Embedded, and Embodied Interaction},
  year = {2024},
  doi = {10.1145/3623509.3633400},
  publisher = {ACM}
}

@inproceedings{lee2022coauthor,
  author = {Lee, Mina and Liang, Percy and Yang, Qian},
  title = {CoAuthor: Designing a Human-AI Collaborative Writing Dataset for Exploring Language Model Capabilities},
  booktitle = {Proceedings of the 2022 CHI Conference on Human Factors in Computing Systems},
  year = {2022},
  doi = {10.1145/3491102.3502030},
  publisher = {ACM}
}

@article{draxler2024ghostwriter,
  author = {Draxler, Fiona and Werner, Anna and Lehmann, Florian and Hoppe, Matthias and Schmidt, Albrecht and Buschek, Daniel and Welsch, Robin},
  title = {The {AI} Ghostwriter Effect: When Users do not Perceive Ownership of {AI}-Generated Text but Self-Declare as Authors},
  journal = {ACM Transactions on Computer-Human Interaction},
  year = {2024},
  doi = {10.1145/3637875}
}

@article{campbell2025effort,
  author = {Campbell, Aidan V. and Wang, Yiyi and Inzlicht, Michael},
  title = {Experimental Evidence that Exerting Effort Increases Meaning},
  journal = {Cognition},
  volume = {257},
  pages = {106065},
  year = {2025},
  doi = {10.1016/j.cognition.2025.106065}
}

@inproceedings{gero2023social,
  author = {Gero, Katy Ilonka and Long, Tao and Chilton, Lydia B},
  title = {Social Dynamics of {AI} Support in Creative Writing},
  booktitle = {Proceedings of the 2023 CHI Conference on Human Factors in Computing Systems},
  year = {2023},
  doi = {10.1145/3544548.3580782},
  publisher = {ACM}
}

@article{rezwana2023cofi,
  author = {Rezwana, Jeba and Maher, Mary Lou},
  title = {Designing Creative {AI} Partners with {COFI}: A Framework for Modeling Interaction in Human-{AI} Co-Creative Systems},
  journal = {ACM Transactions on Computer-Human Interaction},
  year = {2023},
  doi = {10.1145/3519026}
}

@inproceedings{issak2025mosaaic,
  author = {Issak, Alayt and Rezwana, Jeba and Harteveld, Casper},
  title = {{MOSAAIC}: Managing Optimization towards Shared Autonomy, Authority, and Initiative in Co-creation},
  booktitle = {Proceedings of the Sixteenth International Conference on Computational Creativity (ICCC)},
  year = {2025},
  eprint = {2505.11481},
  archivePrefix = {arXiv}
}

@article{singh2025systematic,
  author = {Singh, Saloni and Hindriks, Koen and Heylen, Dirk and Baraka, Kim},
  title = {A Systematic Review of Human-{AI} Co-Creativity},
  journal = {arXiv preprint arXiv:2506.21333},
  year = {2025},
  eprint = {2506.21333},
  archivePrefix = {arXiv}
}

@inproceedings{zhou2026collaposer,
  author = {Zhou, Jiayi and Xie, Liwenhan and Ma, Jiaju and Wei, Zheng and Qu, Huamin and Rao, Anyi},
  title = {{Collaposer}: Transforming Photo Collections into Visual Assets for Storytelling with Collages},
  booktitle = {Proceedings of the 2026 CHI Conference on Human Factors in Computing Systems},
  pages = {1--18},
  year = {2026},
  doi = {10.1145/3772318.3791160},
  publisher = {ACM}
}

@incollection{bodolec2012chinese,
  author    = {Bodolec, Caroline},
  title     = {The {Chinese} Paper-Cut: From Local Inventories to the {UNESCO} Representative List of the Intangible Cultural Heritage of Humanity},
  booktitle = {Heritage Regimes and the State},
  editor    = {Bendix, Regina F. and Eggert, Aditya and Peselmann, Arnika},
  series    = {G{\"o}ttingen Studies in Cultural Property},
  volume    = {6},
  pages     = {249--264},
  publisher = {Universit{\"a}tsverlag G{\"o}ttingen},
  year      = {2012},
  doi       = {10.4000/books.gup.392}
}

@article{gershenfeld2012make,
  author    = {Gershenfeld, Neil},
  title     = {How to Make Almost Anything: The Digital Fabrication Revolution},
  journal   = {Foreign Affairs},
  volume    = {91},
  number    = {6},
  pages     = {43--57},
  year      = {2012}
}

@article{baudisch2017personal,
  author    = {Baudisch, Patrick and Mueller, Stefanie},
  title     = {Personal Fabrication},
  journal   = {Foundations and Trends{\textregistered} in Human--Computer Interaction},
  volume    = {10},
  number    = {3--4},
  pages     = {165--293},
  year      = {2017},
  publisher = {Now Publishers},
  doi       = {10.1561/1100000055}
}

@inproceedings{jacobs2018extending,
  author = {Jacobs, Jennifer and Brandt, Joel and M{\v{e}}ch, Radom{\'\i}r and Resnick, Mitchel},
  title = {Extending Manual Drawing Practices with Artist-Centric Programming Tools},
  booktitle = {Proceedings of the 2018 CHI Conference on Human Factors in Computing Systems},
  pages = {1--12},
  year = {2018},
  publisher = {ACM},
  doi = {10.1145/3173574.3174164}
}

@inproceedings{feng2026cameleon,
  author = {Feng, Shuo and Wang, Xuening and Shan, Yifan and Singh, Krista U. and Liu, Bo and Kwatra, Amritansh and Batra, Ritik and Weinberg, Tobias M. and Roumen, Thijs},
  title = {Comparing Fabrication Workflows in {CAD} to Support Design Reasoning},
  booktitle = {Proceedings of the 2026 CHI Conference on Human Factors in Computing Systems},
  pages = {1--25},
  year = {2026},
  doi = {10.1145/3772318.3790516},
  publisher = {ACM}
}

@inproceedings{devendorf2015being,
  author = {Devendorf, Laura and Ryokai, Kimiko},
  title = {Being the Machine: Reconfiguring Agency and Control in Hybrid Fabrication},
  booktitle = {Proceedings of the 33rd Annual ACM Conference on Human Factors in Computing Systems},
  year = {2015},
  pages = {2477--2486},
  publisher = {ACM},
  doi = {10.1145/2702123.2702547}
}

@book{gershenfeld2005fab,
  author = {Gershenfeld, Neil},
  title = {Fab: The Coming Revolution on Your Desktop---From Personal Computers to Personal Fabrication},
  publisher = {Basic Books},
  year = {2005}
}

@article{zoran2013hybrid,
  author = {Zoran, Amit and Buechley, Leah},
  title = {Hybrid Reassemblage: An Exploration of Craft, Digital Fabrication and Artifact Uniqueness},
  journal = {Leonardo},
  volume = {46},
  number = {1},
  pages = {4--10},
  year = {2013},
  doi = {10.1162/LEON_a_00477}
}

@inproceedings{jain2023vectorfusion,
  author = {Jain, Ajay and Xie, Amber and Abbeel, Pieter},
  title = {{VectorFusion}: Text-to-{SVG} by Abstracting Pixel-Based Diffusion Models},
  booktitle = {Proceedings of the IEEE/CVF Conference on Computer Vision and Pattern Recognition},
  year = {2023},
  pages = {1911--1920},
  doi = {10.1109/CVPR52729.2023.00190}
}

@inproceedings{xing2024svgdreamer,
  author = {Xing, Ximing and Zhou, Haitao and Wang, Chuang and Zhang, Jing and Xu, Dong and Yu, Qian},
  title = {{SVGDreamer}: Text Guided {SVG} Generation with Diffusion Model},
  booktitle = {Proceedings of the IEEE/CVF Conference on Computer Vision and Pattern Recognition},
  year = {2024},
  pages = {4546--4555},
  doi = {10.1109/CVPR52733.2024.00435}
}

@inproceedings{faruqi2025mechstyle,
  author = {Faruqi, Faraz and Abdel-Rahman, Amira and Tejedor, Leandra and Nisser, Martin and Li, Jiaji and Phadnis, Vrushank and Jampani, Varun and Gershenfeld, Neil and Hofmann, Megan and Mueller, Stefanie},
  title = {{MechStyle}: Augmenting Generative {AI} with Mechanical Simulation to Create Stylized and Structurally Viable 3D Models},
  booktitle = {Proceedings of the ACM Symposium on Computational Fabrication},
  pages = {1--15},
  year = {2025},
  doi = {10.1145/3745778.3766655},
  publisher = {ACM}
}

@misc{yin2025qwenimagelayered,
  author = {Yin, Shengming and Zhang, Zekai and Tang, Zecheng and Gao, Kaiyuan and Xu, Xiao and Yan, Kun and Li, Jiahao and Chen, Yilei and Chen, Yuxiang and Shum, Heung-Yeung and Ni, Lionel M. and Zhou, Jingren and Lin, Junyang and Wu, Chenfei},
  title = {Qwen-Image-Layered: Towards Inherent Editability via Layer Decomposition},
  year = {2025},
  eprint = {2512.15603},
  archivePrefix = {arXiv},
  primaryClass = {cs.CV}
}

@inproceedings{li2023beyond,
  title={Beyond the artifact: power as a lens for creativity support tools},
  author={Li, Jingyi and Rawn, Eric and Ritchie, Jacob and Tran O'Leary, Jasper and Follmer, Sean},
  booktitle={Proceedings of the 36th Annual ACM Symposium on User Interface Software and Technology},
  pages={1--15},
  year={2023},
  doi = {10.1145/3586183.3606831}
}

@inproceedings{twigg2021tools,
  author = {Twigg-Smith, Hannah and Tran O'Leary, Jasper and Peek, Nadya},
  title = {Tools, Tricks, and Hacks: Exploring Novel Digital Fabrication Workflows on \#{PlotterTwitter}},
  booktitle = {Proceedings of the 2021 CHI Conference on Human Factors in Computing Systems},
  pages = {1--15},
  year = {2021},
  doi = {10.1145/3411764.3445653},
  publisher = {ACM}
}

@inproceedings{meng2010artistic,
  author = {Meng, Meng and Zhao, Mingtian and Zhu, Song-Chun},
  title = {Artistic Paper-Cut of Human Portraits},
  booktitle = {Proceedings of the 18th ACM International Conference on Multimedia},
  pages = {931--934},
  year = {2010},
  doi = {10.1145/1873951.1874116},
  publisher = {ACM}
}

@inproceedings{wang2025harmonycut,
  author = {Wang, Huanchen and Qiu, Tianrun and Li, Jiaping and Lu, Zhicong and Ma, Yuxin},
  title = {{HarmonyCut}: Supporting Creative {Chinese} Paper-cutting Design with Form and Connotation Harmony},
  booktitle = {Proceedings of the 2025 CHI Conference on Human Factors in Computing Systems},
  pages = {1--22},
  year = {2025},
  doi = {10.1145/3706598.3714159},
  publisher = {ACM}
}

@article{BraunClarke2006,
  author = {Braun, Virginia and Clarke, Victoria},
  title = {Using thematic analysis in psychology},
  journal = {Qualitative Research in Psychology},
  volume = {3},
  number = {2},
  pages = {77--101},
  year = {2006},
  doi = {10.1191/1478088706qp063oa}
}

@article{schmidt2022choice,
  author = {Schmidt, Ruth and Chen, Zeya and Paz Soldan, Veronica},
  title = {Choice Posture, Architecture, and Infrastructure: Systemic Behavioral Design for Public Health Policy},
  journal = {She Ji: The Journal of Design, Economics, and Innovation},
  volume = {8},
  number = {4},
  pages = {504--525},
  year = {2022},
  doi = {10.1016/j.sheji.2022.08.002},
  publisher = {Elsevier}
}

@inproceedings{chen2026privacymotiv,
  author = {Chen, Zeya and Wen, Jianing and Yao, Yaxing and Li, Toby Jia-Jun and Li, Tianshi},
  title = {{PrivacyMotiv}: Vulnerability-Centered Persona Journeys for Empathic Privacy Reviews in {UX} Design},
  booktitle = {Proceedings of the 2026 ACM Designing Interactive Systems Conference},
  pages = {555--587},
  year = {2026},
  doi = {10.1145/3800645.3813014},
  publisher = {ACM}
}

@inproceedings{chen2026framing,
  author = {Chen, Zeya and Pino, Zach and Schmidt, Ruth},
  title = {Framing Data Choices: How Pre-Donation Exploration Designs Influence Data Donation Behavior and Decision-Making},
  booktitle = {Proceedings of DRS2026 Edinburgh},
  year = {2026},
  doi = {10.21606/drs.2026.2558},
  publisher = {Design Research Society}
}

@misc{chen2026comparing,
  author = {Chen, Chaoran and Zhang, Zhiping and Chen, Zeya and Xu, Eryue and Yang, Yinuo and Khalilov, Ibrahim and Gebreegziabher, Simret A. and Ye, Yanfang and Xiao, Ziang and Yao, Yaxing and Li, Tianshi and Li, Toby Jia-Jun},
  title = {Comparing Human Oversight Strategies for Computer-Use Agents},
  year = {2026},
  eprint = {2604.04918},
  archivePrefix = {arXiv},
  primaryClass = {cs.HC},
  url = {https://arxiv.org/abs/2604.04918}
}
